\documentclass[preprintnumbers]{revtex4}
\usepackage{amssymb}
\usepackage{amsfonts}
\usepackage{amsmath}
\usepackage{graphicx}
\usepackage{dcolumn}
\usepackage{bm}
\usepackage{hyperref}
\usepackage{fancyhdr}

\input{tcilatex}
\begin{document}

\preprint{}
\title{Atom interferometer phase in the presence of proof mass}
\author{B. Dubetsky}
\affiliation{bdubetsky@gmail.com}
\date{\today }

\begin{abstract}
This is presented the justification of the expression for the atom
interferometer phase in the presence of proof mass used in
http://arxiv.org/abs/1407.7287. Quantum corrections to that expression are
also derived. The corrections allow one to calculate numerically atom
interferometer phase with accuracy 1ppm or better.
\end{abstract}

\maketitle
\preprint{}

\section{Main relations}

For times between the Raman pulses the atomic density matrix evolves as 
\begin{equation}
i\hbar \dot{\rho}=\left[ H,\rho \right] ,  \label{1}
\end{equation}%
where%
\begin{equation}
H=\QDABOVE{1pt}{p^{2}}{2M_{a}}+U\left( \vec{x},t\right)  \label{2}
\end{equation}%
is Hamiltonian, $\vec{p},M_{a}$ and $U$ are, respectively, atomic momentum,
mass and gravity potential. In the Wigner representation%
\begin{equation}
\rho \left( \vec{x},\vec{p},t\right) =\dfrac{1}{\left( 2\pi \hbar \right)
^{3}}\int d\vec{s}\rho \left( \vec{x}+\dfrac{1}{2}\vec{s},\vec{x}-\dfrac{1}{2%
}\vec{s},t\right) \exp \left( -i\vec{p}\cdot \vec{s}/\hbar \right)  \label{3}
\end{equation}%
one finds \cite{c1}, Sec. III

\begin{subequations}
\label{4}
\begin{gather}
\left\{ \partial _{t}+\dfrac{\vec{p}}{M_{a}}\partial _{\vec{x}}-\partial _{%
\vec{x}}U\left( \vec{x},t\right) \partial _{\vec{p}}+Q\right\} \rho \left( 
\vec{x},\vec{p},t\right) =0,  \label{4a} \\
Q=-\left( i\hbar \right) ^{-1}\left[ U\left( \vec{x}+\dfrac{1}{2}i\hbar
\partial _{\vec{p}},t\right) -U\left( \vec{x}-\dfrac{1}{2}i\hbar \partial _{%
\vec{p}},t\right) \right] +\partial _{\vec{x}}U\left( \vec{x},t\right)
\partial _{\vec{p}}.  \label{4b}
\end{gather}%
When the size of the gravity potential 
\end{subequations}
\begin{equation}
L\sim U\left( \vec{x}\right) /\left\vert \partial _{\vec{x}}U\left( \vec{x}%
\right) \right\vert  \label{4.1}
\end{equation}
is sufficiently large so that 
\begin{equation}
\QDABOVE{1pt}{\hbar }{\Delta pL}\ll 1  \label{5}
\end{equation}%
where $\Delta p$ is a size of the momentum dependence of the density matrix,
one expand over the $\partial _{\vec{p}}$ term and gets approximately 
\begin{subequations}
\label{6}
\begin{eqnarray}
Q &\approx &-\dfrac{\hbar ^{2}}{24}\chi _{ikl}^{\prime }\left( \vec{x}%
,t\right) \partial _{\vec{p}_{i}}\partial _{\vec{p}_{k}}\partial _{\vec{p}%
_{l}}  \label{6a} \\
\chi _{ikl}^{\prime }\left( \vec{x},t\right) &=&-\partial _{x_{i}}\partial
_{x_{k}}\partial _{x_{l}}U\left( \vec{x},t\right)  \label{6b}
\end{eqnarray}%
An implicit summation convention of Eq. (\ref{6}) will be used in all
subsequent equations. Repeated indices and symbols appearing on the
right-hand-side (rhs) of an equation are to be summed over, unless they also
appear on the left-hand side (lhs) of that equation. For the relative weight
of $Q-$term one finds 
\end{subequations}
\begin{equation}
\QDABOVE{1pt}{Q}{\partial _{\vec{x}}U\left( \vec{x},t\right) \partial _{\vec{%
p}}}\sim \dfrac{\hbar ^{2}}{24\left( L\Delta p\right) ^{2}}\ll 1,  \label{7}
\end{equation}%
so that one can consider $Q-$term as a small perturbation and accept that 
\begin{equation}
\rho \left( \vec{x},\vec{p},t\right) =\rho _{0}\left( \vec{x},\vec{p}%
,t\right) +\rho _{Q}\left( \vec{x},\vec{p},t\right) ,  \label{8}
\end{equation}%
where $\rho _{0}\left( \vec{x},\vec{p},t\right) $ is unperturbed density
matrix obeying equation%
\begin{equation}
\left\{ \partial _{t}+\dfrac{\vec{p}}{M_{a}}\partial _{\vec{x}}-\partial _{%
\vec{x}}U\left( \vec{x},t\right) \partial _{\vec{p}}\right\} \rho _{0}\left( 
\vec{x},\vec{p},t\right) =0  \label{9}
\end{equation}%
and $\rho _{Q}\left( \vec{x},\vec{p},t\right) $ is a perturbation evolving as%
\begin{equation}
\left\{ \partial _{t}+\dfrac{\vec{p}}{M_{a}}\partial _{\vec{x}}-\partial _{%
\vec{x}}U\left( \vec{x},t\right) \partial _{\vec{p}}\right\} \rho _{Q}\left( 
\vec{x},\vec{p},t\right) =-Q\rho _{0}\left( \vec{x},\vec{p},t\right)
\label{10}
\end{equation}%
We assume that density matrix is known at some preceding time $t^{\prime },$
i.e. Eqs. (\ref{9}, \ref{10}) are subject to initial conditions 
\begin{subequations}
\label{11}
\begin{eqnarray}
\rho _{0}\left( \vec{x},\vec{p},t^{\prime }\right) &=&\rho \left( \vec{x},%
\vec{p},t^{\prime }\right) ,  \label{11a} \\
\rho _{Q}\left( \vec{x},\vec{p},t^{\prime }\right) &=&0.  \label{11b}
\end{eqnarray}%
Solution of the homogeneous Eq. (\ref{9}) is given by \cite{c1}, Sec III 
\end{subequations}
\begin{equation}
\rho _{0}\left( \vec{x},\vec{p},t\right) =\rho _{0}\left( \vec{R}\left( \vec{%
x},\vec{p},t^{\prime },t\right) ,\vec{P}\left( \vec{x},\vec{p},t^{\prime
},t\right) ,t^{\prime }\right) ,  \label{12}
\end{equation}%
where $\left\{ \vec{R}\left( \vec{x},\vec{p},t_{1},t_{2}\right) ,\vec{P}%
\left( \vec{x},\vec{p},t_{1},t_{2}\right) \right\} $ are atomic classical
position and momentum at time $t_{1}$ subject to initial conditions $\left\{ 
\vec{x},\vec{p}\right\} $ at time $t_{2}.$ Atomic classical trajectory,
evidently, obeys the multiplication law, 
\begin{equation}
\left\{ 
\begin{array}{c}
\vec{R} \\ 
\vec{P}%
\end{array}%
\right\} \left( \vec{R}\left( \vec{x},\vec{p},t^{\prime },t^{\prime \prime
}\right) ,\vec{P}\left( \vec{x},\vec{p},t^{\prime },t^{\prime \prime
}\right) ,t,t^{\prime }\right) =\left\{ 
\begin{array}{c}
\vec{R} \\ 
\vec{P}%
\end{array}%
\right\} \left( \vec{x},\vec{p},t,t^{\prime \prime }\right) .  \label{31}
\end{equation}

Consider now solution of the Eq. (\ref{10}). Since $Q-$term is initially
equal 0, solution of the inhomogeneous Eq. (\ref{10}) contains only the
stimulated part consisting of contributions produced at different times $%
t^{\prime }<t^{\prime \prime }<t.$ Since operator inside the curly brackets
in Eq. (\ref{10}) is a full time derivative,%
\begin{equation*}
\QDABOVE{1pt}{d}{dt}=\partial _{t}+\dfrac{\vec{p}}{M_{a}}\partial _{\vec{x}%
}-\partial _{\vec{x}}U\left( \vec{x},t\right) \partial _{\vec{p}}
\end{equation*}%
in the time interval $\left[ t^{\prime \prime },t^{\prime \prime
}+dt^{\prime \prime }\right] $ one produces contributions to the $Q-$term%
\begin{equation}
d\rho _{Q}\left( \vec{x},\vec{p},t^{\prime \prime }+dt^{\prime \prime
}\right) =-Q\rho _{0}\left( \vec{x},\vec{p},t^{\prime \prime }\right)
dt^{\prime \prime }.  \label{13}
\end{equation}%
Later on this contribution evolves freely [i.e. satisfies Eq.(\ref{10})
subject to the initial condition (\ref{13}) at time $t^{\prime \prime
}+dt^{\prime \prime }$] and at time $t$ one finds for this contribution%
\begin{equation}
d\rho _{Q}\left( \vec{x},\vec{p},t\right) =\dfrac{\hbar ^{2}}{24}dt^{\prime
\prime }\left[ \chi _{ikl}^{\prime }\left( \vec{\xi},t^{\prime \prime
}\right) \partial _{\vec{\pi}_{i}}\partial _{\vec{\pi}_{k}}\partial _{\vec{%
\pi}_{l}}\rho _{0}\left( \vec{\xi},\vec{\pi},t^{\prime \prime }\right) %
\right] _{\vec{\xi}=\vec{R}\left( \vec{x},\vec{p},t^{\prime \prime
},t\right) ,\vec{\pi}=\vec{P}\left( \vec{x},\vec{p},t^{\prime \prime
},t\right) }.  \label{14}
\end{equation}%
Total solution is a sum of all these contributions%
\begin{equation}
\rho _{Q}\left( \vec{x},\vec{p},t\right) =\dfrac{\hbar ^{2}}{24}%
\int_{t^{\prime }}^{t}dt^{\prime \prime }\left[ \chi _{ikl}^{\prime }\left( 
\vec{\xi},t^{\prime \prime }\right) \partial _{\vec{\pi}_{i}}\partial _{\vec{%
\pi}_{k}}\partial _{\vec{\pi}_{l}}\rho _{0}\left( \vec{\xi},\vec{\pi}%
,t^{\prime \prime }\right) \right] _{\vec{\xi}=\vec{R}\left( \vec{x},\vec{p}%
,t^{\prime \prime },t\right) ,\vec{\pi}=\vec{P}\left( \vec{x},\vec{p}%
,t^{\prime \prime },t\right) }.  \label{15.1}
\end{equation}%
Substituting here solution (\ref{12}) for $\rho _{0}$ one finds%
\begin{equation}
\rho _{Q}\left( \vec{x},\vec{p},t\right) =\dfrac{\hbar ^{2}}{24}%
\int_{t^{\prime }}^{t}dt^{\prime \prime }\left[ \chi _{ikl}^{\prime }\left( 
\vec{\xi},t^{\prime \prime }\right) \partial _{\vec{\pi}_{i}}\partial _{\vec{%
\pi}_{k}}\partial _{\vec{\pi}_{l}}\rho _{0}\left( \vec{R}\left( \vec{\xi},%
\vec{\pi},t^{\prime },t^{\prime \prime }\right) ,\vec{P}\left( \vec{\xi},%
\vec{\pi},t^{\prime },t^{\prime \prime }\right) ,t^{\prime }\right) \right]
_{\left\{ 
\begin{array}{c}
_{\vec{\xi}} \\ 
_{\vec{\pi}}%
\end{array}%
\right\} =\left\{ 
\begin{array}{c}
_{\vec{R}} \\ 
_{\vec{P}}%
\end{array}%
\right\} \left( \vec{x},\vec{p},t^{\prime \prime },t\right) }  \label{15}
\end{equation}

\section{AI phase}

Consider now an atom cloud launched at time $t_{0},$ and interacting with $%
\dfrac{\pi }{2}-\pi -\dfrac{\pi }{2}$ sequence of Raman pulses applied at
times%
\begin{equation}
\tau =\left\{ t_{0}+t_{1},t_{0}+t_{1}+T,t_{0}+t_{1}+2T\right\} ,  \label{16}
\end{equation}%
where $t_{1}$ is time delay between cloud launch and 1st Raman pulse. $T$ is
time delay between pulses. We assume that Raman pulses produce coherence
between two hyperfine sublevels $g$ and $e$ of the atomic ground state
manifold, and that initially atomic density matrix (\ref{3}) is given by 
\begin{subequations}
\label{17}
\begin{gather}
\rho _{gg}\left( \vec{x},\vec{p},t_{0}\right) =f\left( \vec{x},\vec{p}%
\right) ,  \label{17a} \\
\rho _{eg}\left( \vec{x},\vec{p},t_{0}\right) =\rho _{ee}\left( \vec{x},\vec{%
p},t_{0}\right) =0.  \label{17b}
\end{gather}

One finds, for example in \cite{c1}, that after $\dfrac{\pi }{2}-$pulse
applied at time $t,$ density matrix elements jump to the values 
\end{subequations}
\begin{subequations}
\label{18}
\begin{gather}
\rho _{ee}\left( \vec{x},\vec{p},t+0\right) =\dfrac{1}{2}\left[ \rho
_{ee}\left( \vec{x},\vec{p},t-0\right) +\rho _{gg}\left( \vec{x},\vec{p}%
-\hbar \vec{k},t-0\right) \right] +\func{Re}\left\{ i\exp \left[ -i\left( 
\vec{k}\cdot \vec{x}-\delta _{12}t-\phi \right) \right] \rho _{eg}\left( 
\vec{x},\vec{p}-\dfrac{\hbar \vec{k}}{2},t-0\right) \right\} ,  \label{18a}
\\
\rho _{gg}\left( \vec{x},\vec{p},t+0\right) =\dfrac{1}{2}\left[ \rho
_{ee}\left( \vec{x},\vec{p}+\hbar \vec{k},t-0\right) +\rho _{gg}\left( \vec{x%
},\vec{p},t-0\right) \right] -\func{Re}\left\{ i\exp \left[ -i\left( \vec{k}%
\cdot \vec{x}-\delta _{12}t-\phi \right) \right] \rho _{eg}\left( \vec{x},%
\vec{p}+\dfrac{\hbar \vec{k}}{2},t-0\right) \right\} ,  \label{18b} \\
\rho _{eg}\left( \vec{x},\vec{p},t+0\right) =\dfrac{i}{2}\exp \left[ i\left( 
\vec{k}\cdot \vec{x}-\delta _{12}t-\phi \right) \right] \left[ \rho
_{ee}\left( \vec{x},\vec{p}+\dfrac{\hbar \vec{k}}{2},t-0\right) -\rho
_{gg}\left( \vec{x},\vec{p}-\dfrac{\hbar \vec{k}}{2},t-0\right) \right] 
\notag \\
+\dfrac{1}{2}\left\{ \rho _{eg}\left( \vec{x},\vec{p},t-0\right) +\exp \left[
2i\left( \vec{k}\cdot \vec{x}-\delta _{12}t-\phi \right) \right] \rho
_{ge}\left( \vec{x},\vec{p},t-0\right) \right\} ,  \label{18c}
\end{gather}%
and after $\pi -$pulse, the density matrix elements jump to the values 
\end{subequations}
\begin{subequations}
\label{19}
\begin{gather}
\rho _{ee}\left( \vec{x},\vec{p},t+0\right) =\rho _{gg}\left( \vec{x},\vec{p}%
-\hbar \vec{k},t-0\right) ,  \label{19a} \\
\rho _{gg}\left( \vec{x},\vec{p},t+0\right) =\rho _{ee}\left( \vec{x},\vec{p}%
+\hbar \vec{k},t-0\right) ,  \label{19b} \\
\rho _{eg}\left( \vec{x},\vec{p},t+0\right) =\exp \left[ 2i\left( \vec{k}%
\vec{x}-\delta _{12}t-\phi \right) \right] \rho _{ge}\left( \vec{x},\vec{p}%
,t-0\right) ,  \label{19c}
\end{gather}%
where $\vec{k}$ is effective wave vector, $\delta _{12}$ is detuning between
fields' frequency difference and hyperfine transition frequency, $\phi $ is
phase difference between traveling components of the Raman field. We allow
pulses to have different detunings and phases $\delta _{12}^{\left( i\right)
},\phi _{i}$ ($i=1,2,3$).

Our purpose is to obtain the atomic density matrix of the excited state $%
\rho _{ee}\left( \vec{x},\vec{p},\tau _{3}+0\right) .$ We will achieve this
by applying consequently Eqs. (\ref{12}, \ref{15}) for density matrix
evolution between Raman pulses and before the 1st Raman pulse, and Eqs. (\ref%
{18}, \ref{19}) for density matrix jumps after the pulse.

Before the first pulse, the density matrix becomes 
\end{subequations}
\begin{equation}
\rho _{gg}\left( \vec{x},\vec{p},\tau _{1}-0\right) =f\left( \vec{R}\left( 
\vec{x},\vec{p},t_{0},\tau _{1}\right) ,\vec{P}\left( \vec{x},\vec{p}%
,t_{0},\tau _{1}\right) \right)  \label{20}
\end{equation}%
After this $\QDABOVE{1pt}{\pi }{2}-$pulse 
\begin{subequations}
\label{21}
\begin{eqnarray}
\rho _{ee}\left( \vec{x},\vec{p},\tau _{1}+0\right) &=&\QDABOVE{1pt}{1}{2}%
f\left( \vec{R}\left( \vec{x},\vec{p}-\hbar \vec{k},t_{0},\tau _{1}\right) ,%
\vec{P}\left( \vec{x},\vec{p}-\hbar \vec{k},t_{0},\tau _{1}\right) \right) ,
\label{21a} \\
\rho _{gg}\left( \vec{x},\vec{p},\tau _{1}+0\right) &=&\QDABOVE{1pt}{1}{2}%
f\left( \vec{R}\left( \vec{x},\vec{p},t_{0},\tau _{1}\right) ,\vec{P}\left( 
\vec{x},\vec{p},t_{0},\tau _{1}\right) \right) ,  \label{21b} \\
\rho _{eg}\left( \vec{x},\vec{p},\tau _{1}+0\right) &=&-\dfrac{i}{2}\exp %
\left[ i\left( \vec{k}\cdot \vec{x}-\delta _{12}^{\left( 1\right) }\tau
_{1}-\phi _{1}\right) \right] f\left( \vec{R}\left( \vec{x},\vec{p}-\QDABOVE{%
1pt}{\hbar \vec{k}}{2},t_{0},\tau _{1}\right) ,\vec{P}\left( \vec{x},\vec{p}-%
\QDABOVE{1pt}{\hbar \vec{k}}{2},t_{0},\tau _{1}\right) \right) .  \label{21c}
\end{eqnarray}%
One uses this density matrix as an initial value at moment $\tau _{1}$ for
the free evolution between 1st and 2nd pulses of the unperturbed density
matrix 
\end{subequations}
\begin{equation}
\rho _{0}\left( \vec{x},\vec{p},\tau _{1}+0\right) =\rho \left( \vec{x},\vec{%
p},\tau _{1}+0\right)  \label{21.1}
\end{equation}

We now consider $Q-$term (\ref{15}) before the second pulse action. From
Eqs. (\ref{15}, \ref{21c}, \ref{21.1}) one obtains 
\begin{gather}
\rho _{Qeg}\left( \vec{x},\vec{p},\tau _{2}-0\right) =-i\dfrac{\hbar ^{2}}{48%
}\int_{\tau _{1}}^{\tau _{2}}dt  \notag \\
\times \left\{ \chi _{ikl}^{\prime }\left( \vec{\xi},t\right) \partial _{%
\vec{\pi}_{i}}\partial _{\vec{\pi}_{k}}\partial _{\vec{\pi}_{l}}\left[ 
\begin{array}{c}
\exp \left[ i\left( \vec{k}\cdot \vec{R}\left( \vec{\xi},\vec{\pi},\tau
_{1},t\right) -\delta _{12}^{\left( 1\right) }\tau _{1}-\phi _{1}\right) %
\right] \\ 
\times f\left( 
\begin{array}{c}
\vec{R}\left( \vec{R}\left( \vec{\xi},\vec{\pi},\tau _{1},t\right) ,\vec{P}%
\left( \vec{\xi},\vec{\pi},\tau _{1},t\right) -\QDABOVE{1pt}{\hbar \vec{k}}{2%
},t_{0},\tau _{1}\right) , \\ 
\vec{P}\left( \vec{R}\left( \vec{\xi},\vec{\pi},\tau _{1},t\right) ,\vec{P}%
\left( \vec{\xi},\vec{\pi},\tau _{1},t\right) -\QDABOVE{1pt}{\hbar \vec{k}}{2%
},t_{0},\tau _{1}\right)%
\end{array}%
\right)%
\end{array}%
\right] \right\} _{\left\{ 
\begin{array}{c}
_{\vec{\xi}} \\ 
_{\vec{\pi}}%
\end{array}%
\right\} =\left\{ 
\begin{array}{c}
_{\vec{R}} \\ 
_{\vec{P}}%
\end{array}%
\right\} \left( \vec{x},\vec{p},t,\tau _{2}\right) }  \label{22}
\end{gather}%
In the Eq. (\ref{22}), the derivative on the momentum $\vec{\pi}$ is of the
order of $\Delta p^{-1},$ where $\Delta p$ is the momentum size of the
factor in brackets. This factor has two terms, phase-factor and atom
distribution $f.$ Initially, for $t=\tau _{1}$ phase-factor is $\vec{\pi}-$%
independent (because $\vec{R}\left( \vec{\xi},\vec{\pi},\tau _{1},\tau
_{1}\right) =\vec{\xi}$) and derivative is of the order of%
\begin{equation}
\partial _{\vec{\pi}_{i}\text{Thermal}}\sim p_{0}^{-1},  \label{23}
\end{equation}%
where%
\begin{equation}
p_{0}=\left( 2M_{a}k_{B}T_{c}\right) ^{1/2}  \label{24}
\end{equation}%
is thermal momentum, the $T_{c}$ is cloud temperature.

Consider now the phase factor $\exp \left[ i\left( \vec{k}\cdot \vec{R}%
\left( \vec{\xi},\vec{\pi},\tau _{1},t\right) -\delta _{12}^{\left( 1\right)
}\tau _{1}-\phi _{1}\right) \right] $ at $t-\tau _{1}\sim T.$ For the
purpose of the estimates lets "turn off" the gravity field. Then atom
trajectory evidently equal to%
\begin{equation}
\vec{R}\left( \vec{\xi},\vec{\pi},\tau _{1},t\right) =\vec{\xi}-\left(
t-\tau _{1}\right) \QDABOVE{1pt}{\vec{\pi}}{M_{a}}  \label{25}
\end{equation}%
and the phase factor acquires Doppler phase $\vec{k}\cdot \QDABOVE{1pt}{\vec{%
\pi}}{M_{a}}\left( t-\tau _{1}\right) .$ It becomes a rapidly oscillating
function of momentum $\vec{\pi}$ having period of the order of%
\begin{equation}
p_{D}\sim M_{a}/kT  \label{26}
\end{equation}%
and phase factor derivative is of magnitude 
\begin{equation}
\partial _{\vec{\pi}_{i}\text{Doppler}}\sim p_{D}^{-1}.  \label{27}
\end{equation}%
For the $^{133}$Cs at temperature $T_{C}\approx 3\mu K,$ $k=1.4743261770\ast
10^{7}$m$^{-1},T=160$ms one finds%
\begin{equation}
\QDABOVE{1pt}{\partial _{\vec{\pi}_{i}\text{Thermal}}}{\partial _{\vec{\pi}%
_{i}\text{Doppler}}}\sim \QDABOVE{1pt}{1}{kv_{0}T}\approx 2\ast 10^{-5}\ll 1,
\label{28}
\end{equation}%
where $v_{0}=p_{0}/M_{a}$ is thermal velocity. Condition (\ref{28}) means
that time separation between pulses $T$ is sufficiently large to be in the
Doppler limiting case, when Doppler phase $kv_{0}T\gg 1.$ In this limit
atomic distribution over momentum is sufficiently smooth to neglect its
derivation, and throughout the text we include only contribution to the $Q-$%
term arising from the derivative of phase factor. For this reason we did not
consider above the $Q-$term for cloud evolution before 1st pulse. As we will
show, the atomic levels' populations ($\rho _{ee}$ and $\rho _{gg}$) have no
phase factor at $t_{0}<t<\tau _{3},$ and therefore $Q-$term arises only from
atomic coherence $\rho _{eg}.$ \textbf{Using the Doppler limiting case (\ref%
{28}) results in AI phase pretty much independent of the atomic momentum and
spatial distribution.}

Calculating derivatives, one obtains%
\begin{gather}
\rho _{Qeg}\left( \vec{x},\vec{p},\tau _{2}-0\right) =-\dfrac{\hbar ^{2}}{48}%
\int_{\tau _{1}}^{\tau _{2}}dt  \notag \\
\times \left[ \vec{k}_{u}\vec{k}_{v}\vec{k}_{w}\chi _{ikl}^{\prime }\left( 
\vec{\xi},t\right) \partial _{\vec{\pi}_{i}}\vec{R}_{u}\left( \vec{\xi},\vec{%
\pi},\tau _{1},t\right) \partial _{\vec{\pi}_{k}}\vec{R}_{v}\left( \vec{\xi},%
\vec{\pi},\tau _{1},t\right) \partial _{\vec{\pi}_{l}}\vec{R}_{w}\left( \vec{%
\xi},\vec{\pi},\tau _{1},t\right) \right] _{\left\{ 
\begin{array}{c}
_{\vec{\xi}} \\ 
_{\vec{\pi}}%
\end{array}%
\right\} =\left\{ 
\begin{array}{c}
_{\vec{R}} \\ 
_{\vec{P}}%
\end{array}%
\right\} \left( \vec{x},\vec{p},t,\tau _{2}\right) }  \notag \\
\left\{ 
\begin{array}{c}
\exp \left[ i\left( \vec{k}\cdot \vec{R}\left( \vec{\xi},\vec{\pi},\tau
_{1},t\right) -\delta _{12}^{\left( 1\right) }\tau _{1}-\phi _{1}\right) %
\right] \\ 
\times f\left( 
\begin{array}{c}
\vec{R}\left( \vec{R}\left( \vec{\xi},\vec{\pi},\tau _{1},t\right) ,\vec{P}%
\left( \vec{\xi},\vec{\pi},\tau _{1},t\right) -\QDABOVE{1pt}{\hbar \vec{k}}{2%
},t_{0},\tau _{1}\right) , \\ 
\vec{P}\left( \vec{R}\left( \vec{\xi},\vec{\pi},\tau _{1},t\right) ,\vec{P}%
\left( \vec{\xi},\vec{\pi},\tau _{1},t\right) -\QDABOVE{1pt}{\hbar \vec{k}}{2%
},t_{0},\tau _{1}\right)%
\end{array}%
\right)%
\end{array}%
\right\} _{\left\{ 
\begin{array}{c}
_{\vec{\xi}} \\ 
_{\vec{\pi}}%
\end{array}%
\right\} =\left\{ 
\begin{array}{c}
_{\vec{R}} \\ 
_{\vec{P}}%
\end{array}%
\right\} \left( \vec{x},\vec{p},t,\tau _{2}\right) }  \label{30}
\end{gather}%
In the described approximation, the derivative no longer acts on the term
inside the curly brackets. Applying the multiplication law \ref{31}, we find
that%
\begin{equation}
\left\{ 
\begin{array}{c}
\vec{R} \\ 
\vec{P}%
\end{array}%
\right\} \left( \vec{\xi},\vec{\pi},\tau _{1},t\right) _{\vec{\xi}=\vec{R}%
\left( \vec{x},\vec{p},t,\tau _{2}\right) ,\vec{\pi}=\vec{P}\left( \vec{x},%
\vec{p},t,\tau _{2}\right) }=\left\{ 
\begin{array}{c}
\vec{R} \\ 
\vec{P}%
\end{array}%
\right\} \left( \vec{x},\vec{p},\tau _{1},\tau _{2}\right) .  \label{32}
\end{equation}%
The expression inside the curly brackets of Eq. (\ref{30}) becomes $t-$%
independent and the $Q-$term in front of the 2nd pulse is then given by%
\begin{gather}
\rho _{Qeg}\left( \vec{x},\vec{p},\tau _{2}-0\right) =-\dfrac{\hbar ^{2}}{48}%
\left\{ \exp \left[ i\left( \vec{k}\cdot \vec{\xi}-\delta _{12}^{\left(
1\right) }\tau _{1}-\phi _{1}\right) \right] f\left( \vec{\xi},\vec{\pi}%
\right) \right\} _{\left\{ 
\begin{array}{c}
_{\vec{\xi}} \\ 
_{\vec{\pi}}%
\end{array}%
\right\} =\left\{ 
\begin{array}{c}
_{\vec{R}\left( \vec{x},\vec{p},\tau _{1},\tau _{2}\right) } \\ 
_{\vec{P}\left( \vec{x},\vec{p},\tau _{1},\tau _{2}\right) -\hbar \vec{k}/2}%
\end{array}%
\right\} }  \notag \\
\times \int_{\tau _{1}}^{\tau _{2}}dt\left[ \vec{k}_{u}\vec{k}_{v}\vec{k}%
_{w}\chi _{ikl}^{\prime }\left( \vec{\xi},t\right) \partial _{\vec{\pi}_{i}}%
\vec{R}_{u}\left( \vec{\xi},\vec{\pi},\tau _{1},t\right) \partial _{\vec{\pi}%
_{k}}\vec{R}_{v}\left( \vec{\xi},\vec{\pi},\tau _{1},t\right) \partial _{%
\vec{\pi}_{l}}\vec{R}_{w}\left( \vec{\xi},\vec{\pi},\tau _{1},t\right) %
\right] _{\left\{ 
\begin{array}{c}
_{\vec{\xi}} \\ 
_{\vec{\pi}}%
\end{array}%
\right\} =\left\{ 
\begin{array}{c}
_{\vec{R}} \\ 
_{\vec{P}}%
\end{array}%
\right\} \left( \vec{x},\vec{p},t,\tau _{2}\right) }  \label{33}
\end{gather}

From Eqs. (\ref{12}, \ref{21}, \ref{31}) 
\begin{subequations}
\label{34}
\begin{gather}
\rho _{ee}\left( \vec{x},\vec{p},\tau _{2}-0\right) =\QDABOVE{1pt}{1}{2}%
f\left( \vec{R}\left( \vec{\xi},\vec{\pi},t_{0},\tau _{1}\right) ,\vec{P}%
\left( \vec{\xi},\vec{\pi},t_{0},\tau _{1}\right) \right) _{\vec{\xi}=\vec{R}%
\left( \vec{x},\vec{p},\tau _{1},\tau _{2}\right) ,\vec{\pi}=\vec{P}\left( 
\vec{x},\vec{p},\tau _{1},\tau _{2}\right) -\hbar \vec{k}},  \label{34a} \\
\rho _{gg}\left( \vec{x},\vec{p},\tau _{2}-0\right) =\QDABOVE{1pt}{1}{2}%
f\left( \vec{R}\left( \vec{x},\vec{p},t_{0},\tau _{2}\right) ,\vec{P}\left( 
\vec{x},\vec{p},t_{0},\tau _{2}\right) \right) ,  \label{34b} \\
\rho _{0eg}\left( \vec{x},\vec{p},\tau _{2}-0\right) =-\dfrac{i}{2}\left\{
\exp \left[ i\left( \vec{k}\cdot \vec{\xi}-\delta _{12}^{\left( 1\right)
}\tau _{1}-\phi _{1}\right) \right] \right.  \notag \\
\left. \times f\left( \vec{R}\left( \vec{\xi},\vec{\pi},t_{0},\tau
_{1}\right) ,\vec{P}\left( \vec{\xi},\vec{\pi},t_{0},\tau _{1}\right)
\right) \right\} _{\vec{\xi}=\vec{R}\left( \vec{x},\vec{p},\tau _{1},\tau
_{2}\right) ,\vec{\pi}=\vec{P}\left( \vec{x},\vec{p},\tau _{1},\tau
_{2}\right) -\hbar \vec{k}/2}  \label{34c}
\end{gather}%
From Eqs. (\ref{19}) after the 2nd pulse density matrix jumps to the value

\end{subequations}
\begin{subequations}
\label{35}
\begin{eqnarray}
\rho _{ee}\left( \vec{x},\vec{p},\tau _{2}+0\right) &=&\QDABOVE{1pt}{1}{2}%
f\left( \vec{R}\left( \vec{x},\vec{p}-\hbar \vec{k},t_{0},\tau _{2}\right) ,%
\vec{P}\left( \vec{x},\vec{p}-\hbar \vec{k},t_{0},\tau _{2}\right) \right) ,
\label{35a} \\
\rho _{gg}\left( \vec{x},\vec{p},\tau _{2}+0\right) &=&\QDABOVE{1pt}{1}{2}%
f\left( \vec{R}\left( \vec{\xi},\vec{\pi},t_{0},\tau _{1}\right) ,\vec{P}%
\left( \vec{\xi},\vec{\pi},t_{0},\tau _{1}\right) \right) _{\vec{\xi}=\vec{R}%
\left( \vec{x},\vec{p}+\hbar \vec{k},\tau _{1},\tau _{2}\right) ,\vec{\pi}=%
\vec{P}\left( \vec{x},\vec{p}+\hbar \vec{k},\tau _{1},\tau _{2}\right)
-\hbar \vec{k}}  \label{35b} \\
\rho _{0eg}\left( \vec{x},\vec{p},\tau _{2}+0\right) &=&\dfrac{i}{2}\left\{
\exp \left\{ i\left[ \vec{k}\cdot \left( 2\vec{x}-\vec{\xi}\right) -2\delta
_{12}^{\left( 2\right) }\tau _{2}+\delta _{12}^{\left( 1\right) }\tau
_{1}-2\phi _{2}+\phi _{1}\right] \right\} \right.  \notag \\
&&\left. \times f\left( \vec{R}\left( \vec{\xi},\vec{\pi},t_{0},\tau
_{1}\right) ,\vec{P}\left( \vec{\xi},\vec{\pi},t_{0},\tau _{1}\right)
\right) \right\} _{\vec{\xi}=\vec{R}\left( \vec{x},\vec{p},\tau _{1},\tau
_{2}\right) ,\vec{\pi}=\vec{P}\left( \vec{x},\vec{p},\tau _{1},\tau
_{2}\right) -\hbar \vec{k}/2},  \label{35c} \\
\rho _{Qeg}\left( \vec{x},\vec{p},\tau _{2}+0\right) &=&-\dfrac{\hbar ^{2}}{%
48}\left\{ \exp \left\{ i\left[ \vec{k}\cdot \left( 2\vec{x}-\vec{\xi}%
\right) -2\delta _{12}^{\left( 2\right) }\tau _{2}+\delta _{12}^{\left(
1\right) }\tau _{1}-2\phi _{2}+\phi _{1}\right] \right\} \right.  \notag \\
&&\left. \times f\left( \vec{R}\left( \vec{\xi},\vec{\pi},t_{0},\tau
_{1}\right) ,\vec{P}\left( \vec{\xi},\vec{\pi},t_{0},\tau _{1}\right)
\right) \right\} _{\vec{\xi}=\vec{R}\left( \vec{x},\vec{p},\tau _{1},\tau
_{2}\right) ,\vec{\pi}=\vec{P}\left( \vec{x},\vec{p},\tau _{1},\tau
_{2}\right) -\hbar \vec{k}/2}  \notag \\
&&\times \int_{\tau _{1}}^{\tau _{2}}dt\left[ \vec{k}_{u}\vec{k}_{v}\vec{k}%
_{w}\chi _{ikl}^{\prime }\left( \vec{\xi},t\right) \right.  \notag \\
&&\left. \partial _{\vec{\pi}_{i}}\vec{R}_{u}\left( \vec{\xi},\vec{\pi},\tau
_{1},t\right) \partial _{\vec{\pi}_{k}}\vec{R}_{v}\left( \vec{\xi},\vec{\pi}%
,\tau _{1},t\right) \partial _{\vec{\pi}_{l}}\vec{R}_{w}\left( \vec{\xi},%
\vec{\pi},\tau _{1},t\right) \right] _{\left\{ 
\begin{array}{c}
_{\vec{\xi}} \\ 
_{\vec{\pi}}%
\end{array}%
\right\} =\left\{ 
\begin{array}{c}
_{\vec{R}} \\ 
_{\vec{P}}%
\end{array}%
\right\} \left( \vec{x},\vec{p},t,\tau _{2}\right) }.  \label{35d}
\end{eqnarray}

Consider now $Q-$term produced during free evolution inside the time
interval $\left[ \tau _{2},\tau _{3}\right] .$ Each density matrix element (%
\ref{35}) produces $Q-$term. However, since diagonal matrix elements (\ref%
{35a}, \ref{35b}) contain no rapidly oscillating in momentum space phase
factors and we neglect their $Q-terms$. Term (\ref{35d}) is already linear
in $Q$ term and can produce only higher order non linear in $Q$
contributions, which we are not including yet. So one has to consider only
the $Q-$term produced by coherence (\ref{35c}), which we denote as $\rho
_{Qeg}^{\prime }.$ From Eq. (\ref{15}) one finds

\end{subequations}
\begin{eqnarray}
\rho _{Qeg}^{\prime }\left( \vec{x},\vec{p},\tau _{3}-0\right) &=&i\QDABOVE{%
1pt}{\hbar ^{2}}{48}\int_{\tau _{2}}^{\tau _{3}}dt\left\{ \chi
_{ikl}^{\prime }\left( \vec{\xi},t\right) \partial _{\vec{\pi}_{i}}\partial
_{\vec{\pi}_{k}}\partial _{\vec{\pi}_{l}}\exp \left\{ i\left[ 2\vec{k}\cdot 
\vec{R}\left( \vec{\xi},\vec{\pi},\tau _{2},t\right) \right. \right. \right.
\notag \\
&&-\left. \left. \vec{k}\cdot \vec{R}\left( \vec{\xi},\vec{\pi},\tau
_{1},t\right) -2\delta _{12}^{\left( 2\right) }\tau _{2}+\delta
_{12}^{\left( 1\right) }\tau _{1}-2\phi _{2}+\phi _{1}\right] \right\} 
\notag \\
&&\times f\left( \vec{R}\left( \vec{R}\left( \vec{R}\left( \vec{\xi},\vec{\pi%
},\tau _{2},t\right) ,\vec{P}\left( \vec{\xi},\vec{\pi},\tau _{2},t\right)
,\tau _{1},\tau _{2}\right) ,\right. \right.  \notag \\
&&\left. \vec{P}\left( \vec{R}\left( \vec{\xi},\vec{\pi},\tau _{2},t\right) ,%
\vec{P}\left( \vec{\xi},\vec{\pi},\tau _{2},t\right) ,\tau _{1},\tau
_{2}\right) -\QDABOVE{1pt}{\hbar \vec{k}}{2},t_{0},\tau _{1}\right) ,  \notag
\\
&&P\left( \vec{R}\left( \vec{R}\left( \vec{\xi},\vec{\pi},\tau _{2},t\right)
,\vec{P}\left( \vec{\xi},\vec{\pi},\tau _{2},t\right) ,\tau _{1},\tau
_{2}\right) ,\right.  \notag \\
&&\left. \left. \vec{P}\left( \vec{R}\left( \vec{\xi},\vec{\pi},\tau
_{2},t\right) ,\vec{P}\left( \vec{\xi},\vec{\pi},\tau _{2},t\right) ,\tau
_{1},\tau _{2}\right) -\QDABOVE{1pt}{\hbar \vec{k}}{2},t_{0},\tau
_{1}\right) \right\} _{\left\{ 
\begin{array}{c}
_{\vec{\xi}} \\ 
_{\vec{\pi}}%
\end{array}%
\right\} =\left\{ 
\begin{array}{c}
_{\vec{R}} \\ 
_{\vec{P}}%
\end{array}%
\right\} \left( \vec{x},\vec{p},t,\tau _{3}\right) }  \label{36}
\end{eqnarray}%
where we used the multiplication law (\ref{31}), 
\begin{equation}
\vec{R}\left( \vec{R}\left( \vec{\xi},\vec{\pi},\tau _{2},t\right) ,\vec{P}%
\left( \vec{\xi},\vec{\pi},\tau _{2},t\right) ,\tau _{1},\tau _{2}\right) =%
\vec{R}\left( \vec{\xi},\vec{\pi},\tau _{1},t\right) .  \label{36.1}
\end{equation}%
In Eq. (\ref{36}) we differentiate over momentum $\vec{\pi}$ only the
rapidly oscillating exponent. After differentiation, we apply the
multiplication law two more times to the phase factor and distribution $f$,
namely%
\begin{equation}
\left\{ 
\begin{tabular}{l}
$\vec{R}$ \\ 
$\vec{P}$%
\end{tabular}%
\right\} \left( \vec{\xi},\vec{\pi},\tau _{2},t\right) _{\left\{ 
\begin{array}{c}
_{\vec{\xi}} \\ 
_{\vec{\pi}}%
\end{array}%
\right\} =\left\{ 
\begin{array}{c}
_{\vec{R}} \\ 
_{\vec{P}}%
\end{array}%
\right\} \left( \vec{x},\vec{p},t,\tau _{3}\right) }=\left\{ 
\begin{tabular}{l}
$\vec{R}$ \\ 
$\vec{P}$%
\end{tabular}%
\right\} \left( \vec{x},\vec{p},\tau _{2},\tau _{3}\right) ,  \label{37}
\end{equation}%
and therefore%
\begin{eqnarray}
&&\left\{ 
\begin{tabular}{l}
$\vec{R}$ \\ 
$\vec{P}$%
\end{tabular}%
\right\} \left( \vec{R}\left( \vec{\xi},\vec{\pi},\tau _{2},t\right) ,\vec{P}%
\left( \vec{\xi},\vec{\pi},\tau _{2},t\right) ,\tau _{1},\tau _{2}\right)
_{\left\{ 
\begin{array}{c}
_{\vec{\xi}} \\ 
_{\vec{\pi}}%
\end{array}%
\right\} =\left\{ 
\begin{array}{c}
_{\vec{R}} \\ 
_{\vec{P}}%
\end{array}%
\right\} \left( \vec{x},\vec{p},t,\tau _{3}\right) }  \notag \\
&=&\left\{ 
\begin{tabular}{l}
$\vec{R}$ \\ 
$\vec{P}$%
\end{tabular}%
\right\} \left( \vec{R}\left( \vec{x},\vec{p},\tau _{2},\tau _{3}\right) ,%
\vec{P}\left( \vec{x},\vec{p},\tau _{2},\tau _{3}\right) ,\tau _{1},\tau
_{2}\right) =\left\{ 
\begin{tabular}{l}
$\vec{R}$ \\ 
$\vec{P}$%
\end{tabular}%
\right\} \left( \vec{x},\vec{p},\tau _{1},\tau _{3}\right)  \label{38}
\end{eqnarray}%
to find out that these terms become $t-$independent. As a result one gets
for $Q-$term $\rho _{Qeg}^{\prime }$ before the 3-rd pulse

\begin{eqnarray}
\rho _{Qeg}^{\prime }\left( \vec{x},\vec{p},\tau _{3}-0\right) &=&-\QDABOVE{%
1pt}{\hbar ^{2}}{48}\left\{ \exp \left\{ i\left[ \vec{k}\cdot \left( 2\vec{R}%
\left( \vec{x},\vec{p},\tau _{2},\tau _{3}\right) -\vec{\xi}\right) -2\delta
_{12}^{\left( 2\right) }\tau _{2}+\delta _{12}^{\left( 1\right) }\tau
_{1}-2\phi _{2}+\phi _{1}\right] \right\} \right.  \notag \\
&&\left. \times f\left( \vec{R}\left( \vec{\xi},\vec{\pi},t_{0},\tau
_{1}\right) ,\vec{P}\left( \vec{\xi},\vec{\pi},t_{0},\tau _{1}\right)
\right) \right\} _{\left\{ 
\begin{array}{c}
_{\vec{\xi}} \\ 
_{\vec{\pi}}%
\end{array}%
\right\} =\left\{ 
\begin{array}{c}
_{\vec{R}} \\ 
_{\vec{P}}%
\end{array}%
\right\} \vec{\xi}=\vec{R}\left( \vec{x},\vec{p},\tau _{1},\tau _{3}\right) ,%
\vec{\pi}=\vec{P}\left( \vec{x},\vec{p},\tau _{1},\tau _{3}\right) -\hbar 
\vec{k}/2}  \notag \\
&&\times \int_{\tau _{2}}^{\tau _{3}}dt\left\{ \chi _{ikl}^{\prime }\left( 
\vec{\xi},t\right) \vec{k}_{u}\vec{k}_{v}\vec{k}_{w}\left[ \partial _{\vec{%
\pi}_{i}}\left( \vec{R}_{u}\left( \vec{\xi},\vec{\pi},\tau _{1},t\right) -2%
\vec{R}_{u}\left( \vec{\xi},\vec{\pi},\tau _{2},t\right) \right) \right]
\right.  \notag \\
&&\times \left[ \partial _{\vec{\pi}_{k}}\left( \vec{R}_{v}\left( \vec{\xi},%
\vec{\pi},\tau _{1},t\right) -2\vec{R}_{v}\left( \vec{\xi},\vec{\pi},\tau
_{2},t\right) \right) \right]  \notag \\
&&\times \left. \left[ \partial _{\vec{\pi}_{l}}\left( \vec{R}_{w}\left( 
\vec{\xi},\vec{\pi},\tau _{1},t\right) -2\vec{R}_{w}\left( \vec{\xi},\vec{\pi%
},\tau _{2},t\right) \right) \right] \right\} _{\left\{ 
\begin{array}{c}
_{\vec{\xi}} \\ 
_{\vec{\pi}}%
\end{array}%
\right\} =\left\{ 
\begin{array}{c}
_{\vec{R}} \\ 
_{\vec{P}}%
\end{array}%
\right\} \left( \vec{x},\vec{p},t,\tau _{3}\right) }.  \label{39}
\end{eqnarray}

Applying Eq. (\ref{12}) one finds that other matrix elements (\ref{35}) in
front of the 3rd pulse become

\begin{subequations}
\label{40}
\begin{eqnarray}
\rho _{ee}\left( \vec{x},\vec{p},\tau _{3}-0\right) &=&\QDABOVE{1pt}{1}{2}%
f\left( \vec{R}\left( \vec{\xi},\vec{\pi},t_{0},\tau _{2}\right) ,\vec{P}%
\left( \vec{\xi},\vec{\pi},t_{0},\tau _{2}\right) \right) _{\vec{\xi}=\vec{R}%
\left( \vec{x},\vec{p},\tau _{2},\tau _{3}\right) ,\vec{\pi}=\vec{P}\left( 
\vec{x},\vec{p},\tau _{2},\tau _{3}\right) -\hbar \vec{k}};  \label{40a} \\
\rho _{gg}\left( \vec{x},\vec{p},\tau _{3}-0\right) &=&\QDABOVE{1pt}{1}{2}%
f\left( \vec{R}\left( \vec{\xi},\vec{\pi},t_{0},\tau _{1}\right) ,\vec{P}%
\left( \vec{\xi},\vec{\pi},t_{0},\tau _{1}\right) \right) _{%
\begin{array}{l}
_{\left\{ \vec{\xi}=\vec{R}\left( \vec{R}\left( \vec{x},\vec{p},\tau
_{2},\tau _{3}\right) ,\vec{P}\left( \vec{x},\vec{p},\tau _{2},\tau
_{3}\right) +\hbar \vec{k},\tau _{1},\tau _{2}\right) ,\right. } \\ 
_{\left. \vec{\pi}=\vec{P}\left( \vec{R}\left( \vec{x},\vec{p},\tau
_{2},\tau _{3}\right) ,\vec{P}\left( \vec{x},\vec{p},\tau _{2},\tau
_{3}\right) +\hbar \vec{k},\tau _{1},\tau _{2}\right) -\hbar \vec{k}\right\}
}%
\end{array}%
}  \label{40b} \\
\rho _{0eg}\left( \vec{x},\vec{p},\tau _{3}-0\right) &=&\dfrac{i}{2}\left\{
\exp \left\{ i\left[ 2\vec{k}\vec{R}\left( \vec{x},\vec{p},\tau _{2},\tau
_{3}\right) -\vec{k}\cdot \vec{\xi}-2\delta _{12}^{\left( 2\right) }\tau
_{2}+\delta _{12}^{\left( 1\right) }\tau _{1}-2\phi _{2}+\phi _{1}\right]
\right\} \right.  \notag \\
&&\left. \times f\left( \vec{R}\left( \vec{\xi},\vec{\pi},t_{0},\tau
_{1}\right) ,\vec{P}\left( \vec{\xi},\vec{\pi},t_{0},\tau _{1}\right)
\right) \right\} _{\vec{\xi}=\vec{R}\left( \vec{x},\vec{p},\tau _{1},\tau
_{3}\right) ,\vec{\pi}=\vec{P}\left( \vec{x},\vec{p},\tau _{1},\tau
_{3}\right) -\hbar \vec{k}/2},  \label{40c} \\
\rho _{Qeg}\left( \vec{x},\vec{p},\tau _{3}-0\right) &=&-\dfrac{\hbar ^{2}}{%
48}\left\{ \exp \left\{ i\left[ 2\vec{k}\vec{R}\left( \vec{x},\vec{p},\tau
_{2},\tau _{3}\right) -\vec{k}\cdot \vec{\xi}-2\delta _{12}^{\left( 2\right)
}\tau _{2}+\delta _{12}^{\left( 1\right) }\tau _{1}-2\phi _{2}+\phi _{1}%
\right] \right\} \right.  \notag \\
&&\left. \times f\left( \vec{R}\left( \vec{\xi},\vec{\pi},t_{0},\tau
_{1}\right) ,\vec{P}\left( \vec{\xi},\vec{\pi},t_{0},\tau _{1}\right)
\right) \right\} _{\vec{\xi}=\vec{R}\left( \vec{x},\vec{p},\tau _{1},\tau
_{3}\right) ,\vec{\pi}=\vec{P}\left( \vec{x},\vec{p},\tau _{1},\tau
_{3}\right) -\hbar \vec{k}/2}  \notag \\
&&\times \int_{\tau _{1}}^{\tau _{2}}dt\left[ \vec{k}_{u}\vec{k}_{v}\vec{k}%
_{w}\chi _{ikl}^{\prime }\left( \vec{\xi},t\right) \partial _{\vec{\pi}_{i}}%
\vec{R}_{u}\left( \vec{\xi},\vec{\pi},\tau _{1},t\right) \partial _{\vec{\pi}%
_{k}}\vec{R}_{v}\left( \vec{\xi},\vec{\pi},\tau _{1},t\right) \right.  \notag
\\
&&\times \left. \partial _{\vec{\pi}_{l}}\vec{R}_{w}\left( \vec{\xi},\vec{\pi%
},\tau _{1},t\right) \right] _{\left\{ 
\begin{array}{c}
_{\vec{\xi}} \\ 
_{\vec{\pi}}%
\end{array}%
\right\} =\left\{ 
\begin{array}{c}
_{\vec{R}} \\ 
_{\vec{P}}%
\end{array}%
\right\} \left( \vec{x},\vec{p},t,\tau _{3}\right) }.  \label{40d}
\end{eqnarray}%
Aggregating different parts of the nondiagonal density matrix elements (\ref%
{40c}, \ref{40d}, \ref{39}) one concludes that in the linear in $Q-$term
approximation this term acts only on the atomic coherence phase:

\end{subequations}
\begin{subequations}
\label{41}
\begin{eqnarray}
\rho _{eg}\left( \vec{x},\vec{p},\tau _{3}-0\right) &\approx &\rho
_{0eg}\left( \vec{x},\vec{p},\tau _{3}-0\right) +\rho _{Qeg}\left( \vec{x},%
\vec{p},\tau _{3}-0\right) +\rho _{Qeg}^{\prime }\left( \vec{x},\vec{p},\tau
_{3}-0\right)  \notag \\
&=&\dfrac{i}{2}\left\{ \exp \left\{ i\left[ 2\vec{k}\vec{R}\left( \vec{x},%
\vec{p},\tau _{2},\tau _{3}\right) -\vec{k}\cdot \vec{\xi}-\tilde{\phi}%
_{Q}\left( \vec{x},\vec{p}\right) -2\delta _{12}^{\left( 2\right) }\tau
_{2}+\delta _{12}^{\left( 1\right) }\tau _{1}-2\phi _{2}+\phi _{1}\right]
\right\} \right.  \notag \\
&&\left. \times f\left( \vec{R}\left( \vec{\xi},\vec{\pi},t_{0},\tau
_{1}\right) ,\vec{P}\left( \vec{\xi},\vec{\pi},t_{0},\tau _{1}\right)
\right) \right\} _{\vec{\xi}=\vec{R}\left( \vec{x},\vec{p},\tau _{1},\tau
_{3}\right) ,\vec{\pi}=\vec{P}\left( \vec{x},\vec{p},\tau _{1},\tau
_{3}\right) -\hbar \vec{k}/2},  \label{41a} \\
\tilde{\phi}_{Q}\left( \vec{x},\vec{p}\right) &=&-\dfrac{\hbar ^{2}}{24}\vec{%
k}_{u}\vec{k}_{v}\vec{k}_{w}\left\{ \int_{\tau _{1}}^{\tau _{2}}dt\chi
_{ikl}^{\prime }\left( \vec{\xi},t\right) \partial _{\vec{\pi}_{i}}\vec{R}%
_{u}\left( \vec{\xi},\vec{\pi},\tau _{1},t\right) \partial _{\vec{\pi}_{k}}%
\vec{R}_{v}\left( \vec{\xi},\vec{\pi},\tau _{1},t\right) \partial _{\vec{\pi}%
_{l}}\vec{R}_{w}\left( \vec{\xi},\vec{\pi},\tau _{1},t\right) \right.  \notag
\\
&&+\int_{\tau _{2}}^{\tau _{3}}dt\chi _{ikl}^{\prime }\left( \vec{\xi}%
,t\right) \left[ \partial _{\vec{\pi}_{i}}\left( \vec{R}_{u}\left( \vec{\xi},%
\vec{\pi},\tau _{1},t\right) -2\vec{R}_{u}\left( \vec{\xi},\vec{\pi},\tau
_{2},t\right) \right) \right]  \notag \\
&&\times \left[ \partial _{\vec{\pi}_{k}}\left( \vec{R}_{v}\left( \vec{\xi},%
\vec{\pi},\tau _{1},t\right) -2\vec{R}_{v}\left( \vec{\xi},\vec{\pi},\tau
_{2},t\right) \right) \right]  \notag \\
&&\times \left. \left[ \partial _{\vec{\pi}_{l}}\left( \vec{R}_{w}\left( 
\vec{\xi},\vec{\pi},\tau _{1},t\right) -2\vec{R}_{w}\left( \vec{\xi},\vec{\pi%
},\tau _{2},t\right) \right) \right] \right\} _{\left\{ 
\begin{array}{c}
_{\vec{\xi}} \\ 
_{\vec{\pi}}%
\end{array}%
\right\} =\left\{ 
\begin{array}{c}
_{\vec{R}} \\ 
_{\vec{P}}%
\end{array}%
\right\} \left( \vec{x},\vec{p},t,\tau _{3}\right) }.  \label{41b}
\end{eqnarray}%
For the $\QDABOVE{1pt}{\pi }{2}-\pi -\QDABOVE{1pt}{\pi }{2}$ AI, after the
3rd pulse, one should calculate only the atomic distribution in the excited
sublevel. One finds from the Eqs. (\ref{18a}, \ref{40a}, \ref{40b}, \ref{41})

\end{subequations}
\begin{eqnarray}
\rho _{ee}\left( \vec{x},\vec{p},\tau _{3}+0\right) &=&\QDABOVE{1pt}{1}{4}%
f\left( \vec{R}\left( \vec{\xi},\vec{\pi},t_{0},\tau _{2}\right) ,\vec{P}%
\left( \vec{\xi},\vec{\pi},t_{0},\tau _{2}\right) \right) _{\vec{\xi}=\vec{R}%
\left( \vec{x},\vec{p},\tau _{2},\tau _{3}\right) ,\vec{\pi}=\vec{P}\left( 
\vec{x},\vec{p},\tau _{2},\tau _{3}\right) -\hbar \vec{k}}  \notag \\
&&+\QDABOVE{1pt}{1}{4}f\left( \vec{R}\left( \vec{\xi},\vec{\pi},t_{0},\tau
_{1}\right) ,\vec{P}\left( \vec{\xi},\vec{\pi},t_{0},\tau _{1}\right)
\right) _{\left\{ 
\begin{array}{c}
_{\vec{\xi}} \\ 
_{\vec{\pi}}%
\end{array}%
\right\} =\left\{ 
\begin{array}{l}
_{\vec{R}\left( \vec{R}\left( \vec{x},\vec{p}-\hbar \vec{k},\tau _{2},\tau
_{3}\right) ,\vec{P}\left( \vec{x},\vec{p}-\hbar \vec{k},\tau _{2},\tau
_{3}\right) +\hbar \vec{k},\tau _{1},\tau _{2}\right) } \\ 
_{\vec{P}\left( \vec{R}\left( \vec{x},\vec{p}-\hbar \vec{k},\tau _{2},\tau
_{3}\right) ,\vec{P}\left( \vec{x},\vec{p}-\hbar \vec{k},\tau _{2},\tau
_{3}\right) +\hbar \vec{k},\tau _{1},\tau _{2}\right) -\hbar \vec{k}}%
\end{array}%
\right\} }  \notag \\
&&-\QDABOVE{1pt}{1}{2}\left\{ \cos \left[ \vec{k}\vec{x}-2\vec{k}\vec{R}%
\left( \vec{x},\vec{p}-\QDABOVE{1pt}{\hbar \vec{k}}{2},\tau _{2},\tau
_{3}\right) +\vec{k}\cdot \vec{\xi}+\tilde{\phi}_{Q}\left( \vec{x},\vec{p}-%
\QDABOVE{1pt}{\hbar \vec{k}}{2}\right) \right. \right.  \notag \\
&&\left. -\delta _{12}^{\left( 3\right) }\tau _{3}+2\delta _{12}^{\left(
2\right) }\tau _{2}-\delta _{12}^{\left( 1\right) }\tau _{1}-\phi _{3}+2\phi
_{2}-\phi _{1}\right]  \notag \\
&&\left. \times f\left( \vec{R}\left( \vec{\xi},\vec{\pi},t_{0},\tau
_{1}\right) ,\vec{P}\left( \vec{\xi},\vec{\pi},t_{0},\tau _{1}\right)
\right) \right\} _{\vec{\xi}=\vec{R}\left( \vec{x},\vec{p}-\hbar \vec{k}%
/2,\tau _{1},\tau _{3}\right) ,\vec{\pi}=\vec{P}\left( \vec{x},\vec{p}-\hbar 
\vec{k}/2,\tau _{1},\tau _{3}\right) -\hbar \vec{k}/2}.  \label{42}
\end{eqnarray}

This density matrix should be used to calculate any response associated with
atoms on the excited state. We are using it to get the probability of atom
cloud excitation defined as%
\begin{equation}
w=\int d\vec{x}d\vec{p}\rho _{ee}\left( \vec{x},\vec{p},\tau _{3}+0\right) .
\label{43}
\end{equation}%
First two terms in Eq. (\ref{42}) are responsible for background. Since
phase space stays invariant under the atom free motion and recoil during
interactions with Raman pulses, the background equals to $\QDABOVE{1pt}{1}{2}
$ and 
\begin{equation}
w=\QDABOVE{1pt}{1}{2}\left( 1-\tilde{w}\right) ,  \label{44}
\end{equation}%
where interferometric term is given by%
\begin{eqnarray}
\tilde{w} &=&\int d\vec{x}d\vec{p}\left\{ \cos \left[ \vec{k}\vec{x}-2\vec{k}%
\vec{R}\left( \vec{x},\vec{p}-\QDABOVE{1pt}{\hbar \vec{k}}{2},\tau _{2},\tau
_{3}\right) +\vec{k}\cdot \vec{\xi}+\tilde{\phi}_{Q}\left( \vec{x},\vec{p}-%
\QDABOVE{1pt}{\hbar \vec{k}}{2}\right) -\delta _{12}^{\left( 3\right) }\tau
_{3}+2\delta _{12}^{\left( 2\right) }\tau _{2}-\delta _{12}^{\left( 1\right)
}\tau _{1}-\phi _{3}+2\phi _{2}-\phi _{1}\right] \right.  \notag \\
&&\left. \times f\left( \vec{R}\left( \vec{\xi},\vec{\pi},t_{0},\tau
_{1}\right) ,\vec{P}\left( \vec{\xi},\vec{\pi},t_{0},\tau _{1}\right)
\right) \right\} _{\vec{\xi}=\vec{R}\left( \vec{x},\vec{p}-\hbar \vec{k}%
/2,\tau _{1},\tau _{3}\right) ,\vec{\pi}=\vec{P}\left( \vec{x},\vec{p}-\hbar 
\vec{k}/2,\tau _{1},\tau _{3}\right) -\hbar \vec{k}/2}.  \label{45}
\end{eqnarray}%
Selecting as integration variables atomic initial position and momentum at
the moment of atom launching $t_{0},$ 
\begin{equation}
\left\{ \vec{x}^{\prime },\vec{p}^{\prime }\right\} =\left\{ \vec{R}\left( 
\vec{\xi},\vec{\pi},t_{0},\tau _{1}\right) ,\vec{P}\left( \vec{\xi},\vec{\pi}%
,t_{0},\tau _{1}\right) \right\}  \label{46}
\end{equation}%
one obtains 
\begin{subequations}
\label{47}
\begin{eqnarray}
\left\{ \vec{\xi},\vec{\pi}\right\} &=&\left\{ \vec{R}\left( \vec{x}^{\prime
},\vec{p}^{\prime },\tau _{1},t_{0}\right) ,\vec{P}\left( \vec{x}^{\prime },%
\vec{p}^{\prime },\tau _{1},t_{0}\right) \right\} ,  \label{47a} \\
\left\{ \vec{x},\vec{p}\right\} &=&\left\{ \vec{R}\left( \vec{R}\left( \vec{x%
}^{\prime },\vec{p}^{\prime },\tau _{1},t_{0}\right) ,\vec{P}\left( \vec{x}%
^{\prime },\vec{p}^{\prime },\tau _{1},t_{0}\right) +\hbar \vec{k}/2,\tau
_{3},\tau _{1}\right) ,\right.  \notag \\
&&\left. \vec{P}\left( \vec{R}\left( \vec{x}^{\prime },\vec{p}^{\prime
},\tau _{1},t_{0}\right) ,\vec{P}\left( \vec{x}^{\prime },\vec{p}^{\prime
},\tau _{1},t_{0}\right) +\hbar \vec{k}/2,\tau _{3},\tau _{1}\right) +\hbar 
\vec{k}/2\right\} ,  \label{47b} \\
\left\vert \partial \left\{ \vec{x},\vec{p}\right\} /\partial \left\{ \vec{x}%
^{\prime },\vec{p}^{\prime }\right\} \right\vert &=&1,  \label{47c} \\
\vec{R}\left( \vec{x},\vec{p}-\QDABOVE{1pt}{\hbar \vec{k}}{2},\tau _{2},\tau
_{3}\right) &=&\vec{R}\left( \vec{R}\left( \vec{x}^{\prime },\vec{p}^{\prime
},\tau _{1},t_{0}\right) ,\vec{P}\left( \vec{x}^{\prime },\vec{p}^{\prime
},\tau _{1},t_{0}\right) +\hbar \vec{k}/2,\tau _{2},\tau _{1}\right) .
\label{47d}
\end{eqnarray}%
After that, replacing $\left\{ \vec{x}^{\prime },\vec{p}^{\prime }\right\}
\rightarrow \left\{ \vec{x},\vec{p}\right\} ,$ one obtains 
\end{subequations}
\begin{equation}
\tilde{w}=\int d\vec{x}d\vec{p}\cos \left[ \phi \left( \vec{x},\vec{p}%
\right) -\delta _{12}^{\left( 3\right) }\tau _{3}+2\delta _{12}^{\left(
2\right) }\tau _{2}-\delta _{12}^{\left( 1\right) }\tau _{1}-\phi _{3}+2\phi
_{2}-\phi _{1}\right] f\left( \vec{x},\vec{p}\right) ,  \label{47.1}
\end{equation}%
where the phase of the AI is given by 
\begin{subequations}
\label{49}
\begin{eqnarray}
\phi \left( \vec{x},\vec{p}\right) &=&\phi _{r}\left( \vec{x},\vec{p}\right)
+\phi _{Q}\left( \vec{x},\vec{p}\right) ,  \label{49a} \\
\phi _{r}\left( \vec{x},\vec{p}\right) &=&\vec{k}\cdot \left[ \vec{R}\left( 
\vec{\xi},\vec{\pi},\tau _{3},\tau _{1}\right) -2\vec{R}\left( \vec{\xi},%
\vec{\pi},\tau _{2},\tau _{1}\right) +\vec{\xi}\right] _{\left\{ \vec{\xi}=%
\vec{R}\left( \vec{x},\vec{p},\tau _{1},t_{0}\right) ,\vec{\pi}=\vec{P}%
\left( \vec{x},\vec{p},\tau _{1},t_{0}\right) +\hbar \vec{k}/2\right\} },
\label{49b} \\
\phi _{Q}\left( \vec{x},\vec{p}\right) &=&\tilde{\phi}_{Q}\left[ \vec{R}%
\left( \vec{\xi},\vec{\pi},\tau _{3},\tau _{1}\right) ,\vec{P}\left( \vec{\xi%
},\vec{\pi},\tau _{3},\tau _{1}\right) \right] _{\left\{ \vec{\xi}=\vec{R}%
\left( \vec{x},\vec{p},\tau _{1},t_{0}\right) ,\vec{\pi}=\vec{P}\left( \vec{x%
},\vec{p},\tau _{1},t_{0}\right) +\hbar \vec{k}/2\right\} },  \label{49c}
\end{eqnarray}%
where $\tilde{\phi}_{Q}$ is given by Eq. (\ref{41b}).

Part $\phi _{r}$ includes "classical" part of the phase (erasing in the
limit $\hbar \rightarrow 0)$ and recoil effect during interacion with Raman
pulses. For the rotating spherical Earth this this part had been calculated
in \cite{c1}. We calculate here additions to $\phi _{r}$ caused by proof
mass field.

Part $\phi _{Q}$ is originated from quantum correction $Q$ to the density
matrix in Wigner representation equation in time between pulses. If atom
trajectory is smaler than the size of the gravity field (\ref{4.1}), one can
expand gravity field in the vicinity of atoms launching point. Holding only
gravity-gradient terms in that expansion one gets $\phi _{Q}=0,$ because
tensor (\ref{6b}) dissapears. For this reason we did not consider $Q-$term
in \cite{c1}. Part $\phi _{Q}$ one has to know for precize measurements of
the Newtonian gravitational constant \cite{c2,c3,c4}. Keeping in mind this
application we will calculate $\phi _{Q}$ below.

\subsection{Part $\protect\phi _{r}$}

Propagation functions $\left\{ \vec{R}\left( \vec{x},\vec{p},t,t^{\prime
}\right) ,\vec{P}\left( \vec{x},\vec{p},t,t^{\prime }\right) \right\} $,
i.e. atomic position and momentum at time $t$ subject to initial value $%
\left\{ \vec{x},\vec{p}\right\} $ at moment $t^{\prime },$ evolve as 
\end{subequations}
\begin{subequations}
\label{100}
\begin{eqnarray}
\overset{{\LARGE \dot{\rightarrow}}}{R}\left( \vec{x},\vec{p},t,t^{\prime
}\right) &=&\dfrac{\vec{P}\left( \vec{x},\vec{p},t,t^{\prime }\right) }{M_{a}%
},  \label{100a} \\
\overset{{\LARGE \dot{\rightarrow}}}{P}\left( \vec{x},\vec{p},t,t^{\prime
}\right) &=&M_{a}\left\{ \vec{g}+\delta \vec{g}\left[ \vec{R}\left( \vec{x},%
\vec{p},t,t^{\prime }\right) ,t\right] \right\} ,  \label{100b}
\end{eqnarray}%
where $\vec{g}$ is Earth gravity field, $\delta \vec{g}\left( \vec{x}%
,t\right) $ is proof mass gravity field. We neglect in Eq. (\ref{100}) the
gravity-gradient, centrifugal and Coriolis forces caused by the rotating
Earth. When $\delta \vec{g}\left( \vec{x},t\right) $ is a perturbation,
approximate solutions of the Eq. (\ref{10}) are \cite{c5} 
\end{subequations}
\begin{subequations}
\label{101}
\begin{eqnarray}
\vec{R}\left( \vec{x},\vec{p},t,t^{\prime }\right) &\approx &\vec{R}^{\left(
0\right) }\left( \vec{x},\vec{p},t,t^{\prime }\right) +\delta \vec{R}\left( 
\vec{x},\vec{p},t,t^{\prime }\right) ,  \label{101a} \\
\vec{P}\left( \vec{x},\vec{p},t,t^{\prime }\right) &\approx &\vec{P}^{\left(
0\right) }\left( \vec{x},\vec{p},t,t^{\prime }\right) +\delta \vec{P}\left( 
\vec{x},\vec{p},t,t^{\prime }\right) ,  \label{101b} \\
\vec{R}^{\left( 0\right) }\left( \vec{x},\vec{p},t,t^{\prime }\right) &=&%
\vec{x}+\dfrac{\vec{p}}{M_{a}}\left( t-t^{\prime }\right) +\vec{g}\dfrac{%
\left( t-t^{\prime }\right) ^{2}}{2},  \label{101c} \\
\vec{P}^{\left( 0\right) }\left( \vec{x},\vec{p},t,t^{\prime }\right) &=&%
\vec{p}+M_{a}\vec{g}\left( t-t^{\prime }\right) ;  \label{101d} \\
\delta \vec{R}\left( \vec{x},\vec{p},t,t^{\prime }\right) &=&\int_{t^{\prime
}}^{t}dt^{\prime \prime }\left( t-t^{\prime \prime }\right) \delta \vec{g}%
\left[ \vec{R}^{\left( 0\right) }\left( \vec{x},\vec{p},t^{\prime \prime
},t^{\prime }\right) ,t^{\prime \prime }\right]  \label{101e} \\
\delta \vec{P}\left( \vec{x},\vec{p},t,t^{\prime }\right)
&=&M_{a}\int_{t^{\prime }}^{t}dt^{\prime \prime }\delta \vec{g}\left[ \vec{R}%
^{\left( 0\right) }\left( \vec{x},\vec{p},t^{\prime \prime },t^{\prime
}\right) ,t^{\prime \prime }\right] ;.  \label{101f}
\end{eqnarray}%
Functions $\left\{ \vec{R}^{\left( 0\right) },\vec{P}^{\left( 0\right)
},\delta \vec{R},\delta \vec{P}\right\} $ obey multiplication laws 
\end{subequations}
\begin{subequations}
\label{102}
\begin{eqnarray}
\left\{ 
\begin{array}{c}
\vec{R}^{\left( 0\right) } \\ 
\vec{P}^{\left( 0\right) }%
\end{array}%
\right\} \left( \vec{R}^{\left( 0\right) }\left( \vec{x},\vec{p},t^{\prime
},t^{\prime \prime }\right) ,\vec{P}^{\left( 0\right) }\left( \vec{x},\vec{p}%
,t^{\prime },t^{\prime \prime }\right) ,t,t^{\prime }\right) &=&\left\{ 
\begin{array}{c}
\vec{R}^{\left( 0\right) } \\ 
\vec{P}^{\left( 0\right) }%
\end{array}%
\right\} \left( \vec{x},\vec{p},t,t^{\prime \prime }\right) .  \label{102a}
\\
\left\{ 
\begin{array}{c}
\delta \vec{R} \\ 
\delta \vec{P}%
\end{array}%
\right\} \left( \delta \vec{R}\left( \vec{x},\vec{p},t^{\prime },t^{\prime
\prime }\right) ,\delta \vec{P}\left( \vec{x},\vec{p},t^{\prime },t^{\prime
\prime }\right) ,t,t^{\prime }\right) &=&\left\{ 
\begin{array}{c}
\delta \vec{R} \\ 
\delta \vec{P}%
\end{array}%
\right\} \left( \vec{x},\vec{p},t,t^{\prime \prime }\right) .  \label{102b}
\end{eqnarray}%
To get phase (\ref{49b}) one needs aproximate expression for propagator $%
\vec{R}\left( \vec{R}\left( \vec{x},\vec{p},\tau _{1},t_{0}\right) ,\vec{P}%
\left( \vec{x},\vec{p},\tau _{1},t_{0}\right) +\dfrac{\hbar \vec{k}}{2},\tau
_{s},\tau _{1}\right) .$ Using Eqs. (\ref{101}, \ref{102}) one consequently
finds

\end{subequations}
\begin{gather}
\vec{R}\left( \vec{R}\left( \vec{x},\vec{p},\tau _{1},t_{0}\right) ,\vec{P}%
\left( \vec{x},\vec{p},\tau _{1},t_{0}\right) +\dfrac{\hbar \vec{k}}{2},\tau
_{s},\tau _{1}\right) \approx \vec{R}^{\left( 0\right) }\left( 
\begin{array}{c}
\vec{R}^{\left( 0\right) }\left( \vec{x},\vec{p},\tau _{1},t_{0}\right)
+\delta \vec{R}\left( \vec{x},\vec{p},\tau _{1},t_{0}\right) , \\ 
\vec{P}^{\left( 0\right) }\left( \vec{x},\vec{p},\tau _{1},t_{0}\right)
+\delta \vec{P}\left( \vec{x},\vec{p},\tau _{1},t_{0}\right) +\dfrac{\hbar 
\vec{k}}{2},\tau _{s},\tau _{1}%
\end{array}%
\right)  \notag \\
+\delta \vec{R}\left( \vec{R}^{\left( 0\right) }\left( \vec{x},\vec{p},\tau
_{1},t_{0}\right) ,\vec{P}^{\left( 0\right) }\left( \vec{x},\vec{p},\tau
_{1},t_{0}\right) +\dfrac{\hbar \vec{k}}{2},\tau _{s},\tau _{1}\right) 
\notag \\
=\vec{R}^{\left( 0\right) }\left( \vec{x},\vec{p},\tau _{1},t_{0}\right)
+\delta \vec{R}\left( \vec{x},\vec{p},\tau _{1},t_{0}\right) +  \notag \\
\dfrac{1}{M_{a}}\left[ \vec{P}^{\left( 0\right) }\left( \vec{x},\vec{p},\tau
_{1},t_{0}\right) +M_{a}\int_{t_{0}}^{\tau _{1}}dt^{\prime }\delta \vec{g}%
\left[ \vec{R}^{\left( 0\right) }\left( \vec{x},\vec{p},t^{\prime
},t_{0}\right) ,t^{\prime }\right] +\dfrac{\hbar \vec{k}}{2}\right] \left(
\tau _{s}-\tau _{1}\right) +\dfrac{1}{2}\vec{g}\left( \tau _{s}-\tau
_{1}\right) ^{2}  \notag \\
+\int_{\tau _{1}}^{\tau _{s}}dt^{\prime }\left( \tau _{s}-t^{\prime }\right)
\delta \vec{g}\left[ \vec{R}^{\left( 0\right) }\left( \vec{R}^{\left(
0\right) }\left( \vec{x},\vec{p},\tau _{1},t_{0}\right) ,\vec{P}^{\left(
0\right) }\left( \vec{x},\vec{p},\tau _{1},t_{0}\right) +\dfrac{\hbar \vec{k}%
}{2},t^{\prime },\tau _{1}\right) ,t^{\prime }\right] ,  \label{103}
\end{gather}%
Since from Eqs. (\ref{102}, \ref{101c}),%
\begin{eqnarray}
\vec{R}^{\left( 0\right) }\left( \vec{x},\vec{p},\tau _{s},t_{0}\right) &=&%
\vec{R}^{\left( 0\right) }\left( \vec{R}^{\left( 0\right) }\left( \vec{x},%
\vec{p},\tau _{1},t_{0}\right) ,\vec{P}^{\left( 0\right) }\left( \vec{x},%
\vec{p},\tau _{1},t_{0}\right) ,\tau _{s},\tau _{1}\right)  \notag \\
&=&\vec{R}^{\left( 0\right) }\left( \vec{x},\vec{p},\tau _{1},t_{0}\right) +%
\dfrac{1}{M_{a}}\vec{P}^{\left( 0\right) }\left( \vec{x},\vec{p},\tau
_{1},t_{0}\right) \left( \tau _{s}-\tau _{1}\right) +\dfrac{1}{2}\vec{g}%
\left( \tau _{s}-\tau _{1}\right) ^{2}  \label{104}
\end{eqnarray}%
one rewrites%
\begin{gather}
\vec{R}\left( \vec{R}\left( \vec{x},\vec{p},\tau _{1},t_{0}\right) ,\vec{P}%
\left( \vec{x},\vec{p},\tau _{1},t_{0}\right) +\dfrac{\hbar \vec{k}}{2},\tau
_{s},\tau _{1}\right) \approx \vec{R}^{\left( 0\right) }\left( \vec{x},\vec{p%
},\tau _{s},t_{0}\right) +\dfrac{\hbar \vec{k}}{2M_{a}}\left( \tau _{s}-\tau
_{1}\right)  \notag \\
+\left( \tau _{s}-\tau _{1}\right) \int_{t_{0}}^{\tau _{1}}dt^{\prime \prime
}\delta \vec{g}\left[ \vec{R}^{\left( 0\right) }\left( \vec{x},\vec{p}%
,t^{\prime },t_{0}\right) ,t^{\prime }\right] +\int_{t_{0}}^{\tau
_{1}}dt^{\prime \prime }\left( \tau _{1}-t^{\prime }\right) \delta \vec{g}%
\left[ \vec{R}^{\left( 0\right) }\left( \vec{x},\vec{p},t^{\prime
},t_{0}\right) ,t^{\prime }\right]  \notag \\
+\int_{\tau _{1}}^{\tau _{s}}dt^{\prime }\left( \tau _{s}-t^{\prime }\right)
\delta \vec{g}\left[ \vec{R}^{\left( 0\right) }\left( \vec{R}^{\left(
0\right) }\left( \vec{x},\vec{p},\tau _{1},t_{0}\right) ,\vec{P}^{\left(
0\right) }\left( \vec{x},\vec{p},\tau _{1},t_{0}\right) +\dfrac{\hbar \vec{k}%
}{2},t^{\prime },\tau _{1}\right) ,t^{\prime }\right] .  \label{105}
\end{gather}%
One uses now that%
\begin{gather}
\vec{R}^{\left( 0\right) }\left( \vec{R}^{\left( 0\right) }\left( \vec{x},%
\vec{p},\tau _{1},t_{0}\right) ,\vec{P}^{\left( 0\right) }\left( \vec{x},%
\vec{p},\tau _{1},t_{0}\right) +\dfrac{\hbar \vec{k}}{2},t^{\prime },\tau
_{1}\right) =\dfrac{\hbar \vec{k}}{2M_{a}}\left( t^{\prime }-\tau _{1}\right)
\notag \\
+\vec{R}^{\left( 0\right) }\left( \vec{R}^{\left( 0\right) }\left( \vec{x},%
\vec{p},\tau _{1},t_{0}\right) ,\vec{P}^{\left( 0\right) }\left( \vec{x},%
\vec{p},\tau _{1},t_{0}\right) ,t^{\prime },\tau _{1}\right)  \notag \\
=\vec{R}^{\left( 0\right) }\left( \vec{x},\vec{p},t^{\prime },t_{0}\right) +%
\dfrac{\hbar \vec{k}}{2M_{a}}\left( t^{\prime }-\tau _{1}\right) ,
\label{106}
\end{gather}%
and therefore%
\begin{gather}
\vec{R}\left( \vec{R}\left( \vec{x},\vec{p},\tau _{1},t_{0}\right) ,\vec{P}%
\left( \vec{x},\vec{p},\tau _{1},t_{0}\right) +\dfrac{\hbar \vec{k}}{2},\tau
_{s},\tau _{1}\right) \approx \vec{R}^{\left( 0\right) }\left( \vec{x},\vec{p%
},\tau _{s},t_{0}\right) +\dfrac{\hbar \vec{k}}{2M_{a}}\left( \tau _{s}-\tau
_{1}\right)  \notag \\
+\left( \tau _{s}-\tau _{1}\right) \int_{t_{0}}^{\tau _{1}}dt^{\prime \prime
}\delta \vec{g}\left[ \vec{R}^{\left( 0\right) }\left( \vec{x},\vec{p}%
,t^{\prime },t_{0}\right) ,t^{\prime }\right] +\int_{t_{0}}^{\tau
_{1}}dt^{\prime \prime }\left( \tau _{1}-t^{\prime }\right) \delta \vec{g}%
\left[ \vec{R}^{\left( 0\right) }\left( \vec{x},\vec{p},t^{\prime
},t_{0}\right) ,t^{\prime }\right]  \notag \\
+\int_{\tau _{1}}^{\tau _{s}}dt^{\prime }\left( \tau _{s}-t^{\prime }\right)
\delta \vec{g}\left[ \vec{R}^{\left( 0\right) }\left( \vec{x},\vec{p}%
,t^{\prime },t_{0}\right) ,t^{\prime }\right]  \notag \\
+\int_{\tau _{1}}^{\tau _{s}}dt^{\prime }\left( \tau _{s}-t^{\prime }\right)
\left\{ \delta \vec{g}\left[ \vec{R}^{\left( 0\right) }\left( \vec{x},\vec{p}%
,t^{\prime },t_{0}\right) +\dfrac{\hbar \vec{k}}{2M_{a}}\left( t^{\prime
}-\tau _{1}\right) ,t^{\prime }\right] -\delta \vec{g}\left[ \vec{R}^{\left(
0\right) }\left( \vec{x},\vec{p},t^{\prime },t_{0}\right) ,t^{\prime }\right]
\right\} .  \label{107}
\end{gather}%
The propagator (\ref{101e}) can be rewritten as%
\begin{gather}
\delta \vec{R}\left( \vec{x},\vec{p},\tau _{s},t_{0}\right) \equiv
\int_{t_{0}}^{\tau _{s}}dt^{\prime }\left( \tau _{s}-t^{\prime }\right)
\delta \vec{g}\left[ \vec{R}^{\left( 0\right) }\left( \vec{x},\vec{p}%
,t^{\prime },t_{0}\right) ,t^{\prime }\right]  \notag \\
=\int_{\tau _{1}}^{\tau _{s}}dt^{\prime }\left( \tau _{s}-t^{\prime }\right)
\delta \vec{g}\left[ \vec{R}^{\left( 0\right) }\left( \vec{x},\vec{p}%
,t^{\prime },t_{0}\right) ,t^{\prime }\right] +\int_{t_{0}}^{\tau
_{1}}dt^{\prime }\left( \tau _{s}-t^{\prime }\right) \delta \vec{g}\left[ 
\vec{R}^{\left( 0\right) }\left( \vec{x},\vec{p},t^{\prime },t_{0}\right)
,t^{\prime }\right]  \notag \\
=\int_{\tau _{1}}^{\tau _{s}}dt^{\prime }\left( \tau _{s}-t^{\prime }\right)
\delta \vec{g}\left[ \vec{R}^{\left( 0\right) }\left( \vec{x},\vec{p}%
,t^{\prime },t_{0}\right) ,t^{\prime }\right] +\int_{t_{0}}^{\tau
_{1}}dt^{\prime }\left( \tau _{1}-t^{\prime }\right) \delta \vec{g}\left[ 
\vec{R}^{\left( 0\right) }\left( \vec{x},\vec{p},t^{\prime },t_{0}\right)
,t^{\prime }\right]  \notag \\
+\left( \tau _{s}-\tau _{1}\right) \int_{t_{0}}^{\tau _{1}}dt^{\prime
}\delta \vec{g}\left[ \vec{R}^{\left( 0\right) }\left( \vec{x},\vec{p}%
,t^{\prime },t_{0}\right) ,t^{\prime }\right] ,  \label{108}
\end{gather}%
which coincides with the sum of 3rd, 4th and 5th terms on the
right-hand-side (rhs) of Eq. (\ref{107}). Finally 
\begin{gather}
\vec{R}\left( \vec{R}\left( \vec{x},\vec{p},\tau _{1},t_{0}\right) ,\vec{P}%
\left( \vec{x},\vec{p},\tau _{1},t_{0}\right) +\dfrac{\hbar \vec{k}}{2},\tau
_{s},\tau _{1}\right) \approx \vec{R}^{\left( 0\right) }\left( \vec{x},\vec{p%
},\tau _{s},t_{0}\right) +\delta \vec{R}\left( \vec{x},\vec{p},\tau
_{s},t_{0}\right) +\dfrac{\hbar \vec{k}}{2M_{a}}\left( \tau _{s}-\tau
_{1}\right)  \notag \\
+\int_{\tau _{1}}^{\tau _{s}}dt^{\prime }\left( \tau _{s}-t^{\prime }\right)
\left\{ \delta \vec{g}\left[ \vec{R}^{\left( 0\right) }\left( \vec{x},\vec{p}%
,t^{\prime },t_{0}\right) +\dfrac{\hbar \vec{k}}{2M_{a}}\left( t^{\prime
}-\tau _{1}\right) ,t^{\prime }\right] -\delta \vec{g}\left[ \vec{R}^{\left(
0\right) }\left( \vec{x},\vec{p},t^{\prime },t_{0}\right) ,t^{\prime }\right]
\right\} .  \label{200}
\end{gather}%
Substituting this result in the brackets of Eq. (\ref{49b}) for the 1st $%
\left( s=3\right) $ and 2nd $\left( s=2\right) $ terms, one finds that the
phase of the atom interferometer consist of the terms corresponding to Earth
gravity and the proof mass gravity field, and quantum correction, 
\begin{subequations}
\label{201}
\begin{eqnarray}
\phi _{r}\left( \vec{x},\vec{p}\right) &=&\phi _{0}\left( \vec{x},\vec{p}%
\right) +\delta \phi \left( \vec{x},\vec{p}\right) +\phi _{q}\left( \vec{x},%
\vec{p}\right) ,  \label{201a} \\
\phi _{0}\left( \vec{x},\vec{p}\right) &=&\vec{R}^{\left( 0\right) }\left( 
\vec{x},\vec{p},\tau _{3},t_{0}\right) -2\vec{R}^{\left( 0\right) }\left( 
\vec{x},\vec{p},\tau _{2},t_{0}\right) +\vec{R}^{\left( 0\right) }\left( 
\vec{x},\vec{p},\tau _{1},t_{0}\right) =\vec{k}\vec{g}T^{2},  \label{201b} \\
\delta \phi \left( \vec{x},\vec{p}\right) &=&\vec{k}\cdot \left[ \delta \vec{%
R}\left( \vec{x},\vec{p},\tau _{3},t_{0}\right) -2\delta \vec{R}\left( \vec{x%
},\vec{p},\tau _{2},t_{0}\right) +\delta \vec{R}\left( \vec{x},\vec{p},\tau
_{1},t_{0}\right) \right] ,  \label{201c} \\
\phi _{q}\left( \vec{x},\vec{p}\right) &=&\vec{k}\cdot \vec{\psi}_{q},
\label{201d} \\
\vec{\psi}_{q} &=&\int_{\tau 1}^{\tau _{3}}dt^{\prime }\left( \tau
_{3}-t^{\prime }\right) \left\{ \delta \vec{g}\left[ \vec{R}^{\left(
0\right) }\left( \vec{x},\vec{p},t^{\prime },t_{0}\right) +\dfrac{\hbar \vec{%
k}}{2M_{a}}\left( t^{\prime }-\tau _{1}\right) ,t^{\prime }\right] -\delta 
\vec{g}\left[ \vec{R}^{\left( 0\right) }\left( \vec{x},\vec{p},t^{\prime
},t_{0}\right) ,t^{\prime }\right] \right\}  \notag \\
&&-2\int_{\tau _{1}}^{\tau _{2}}\left( \tau _{2}-t^{\prime }\right) \left\{
\delta \vec{g}\left[ \vec{R}^{\left( 0\right) }\left( \vec{x},\vec{p}%
,t^{\prime },t_{0}\right) +\dfrac{\hbar \vec{k}}{2M_{a}}\left( t^{\prime
}-\tau _{1}\right) ,t^{\prime }\right] -\delta \vec{g}\left[ \vec{R}^{\left(
0\right) }\left( \vec{x},\vec{p},t^{\prime },t_{0}\right) ,t^{\prime }\right]
\right\} .  \label{201e}
\end{eqnarray}%
Using Eq. (\ref{101e}) and piecinmg together terms corresponding to the
integrations in time intervals $\left[ t_{0},\tau _{1}\right] ,$ $\left[
\tau _{1},\tau _{2}\right] ,$ and $\left[ \tau _{2},\tau _{3}\right] $ one
gets \cite{c6} 
\end{subequations}
\begin{subequations}
\label{202}
\begin{align}
\delta \phi & =\vec{k}\cdot \left( \tau _{3}\vec{u}_{30}-t_{1}\vec{u}_{20}+%
\vec{u}_{21}-\vec{u}_{31}\right) ,  \label{202a} \\
\vec{u}_{\alpha \beta }& =\int_{\tau _{\alpha -1}}^{\tau _{\alpha
}}dtt^{\beta }\delta \vec{g}\left( \vec{a}+\vec{v}t+\vec{g}_{E}\dfrac{t^{2}}{%
2},t\right) ,  \label{202b}
\end{align}

Consider now quantum correction (\ref{201d}), vector (\ref{201e}) can be
rewritten as 
\end{subequations}
\begin{eqnarray}
\vec{\psi}_{q} &=&\int_{\tau _{2}}^{\tau _{3}}dt^{\prime }\left( \tau
_{3}-t^{\prime }\right) \left\{ \delta \vec{g}\left[ \vec{R}^{\left(
0\right) }\left( \vec{x},\vec{p},t^{\prime },t_{0}\right) +\dfrac{\hbar \vec{%
k}}{2M_{a}}\left( t^{\prime }-\tau _{1}\right) ,t^{\prime }\right] -\delta 
\vec{g}\left[ \vec{R}^{\left( 0\right) }\left( \vec{x},\vec{p},t^{\prime
},t_{0}\right) ,t^{\prime }\right] \right\}  \notag \\
&&+\int_{\tau _{1}}^{\tau _{2}}dt^{\prime }\left( t^{\prime }-\tau
_{1}\right) \left\{ \delta \vec{g}\left[ \vec{R}^{\left( 0\right) }\left( 
\vec{x},\vec{p},t^{\prime },t_{0}\right) +\dfrac{\hbar \vec{k}}{2M_{a}}%
\left( t^{\prime }-\tau _{1}\right) ,t^{\prime }\right] -\delta \vec{g}\left[
\vec{R}^{\left( 0\right) }\left( \vec{x},\vec{p},t^{\prime },t_{0}\right)
,t^{\prime }\right] \right\} .  \label{203}
\end{eqnarray}%
Substituting $t^{\prime }=\tau _{2}+\theta $ for the 1st term of Eq. (\ref%
{23}), and $t^{\prime }=\tau _{1}+\theta $ for the 2nd term one finds%
\begin{eqnarray}
\vec{\psi}_{q} &=&\int_{0}^{T}d\theta \left\{ \left( T-\theta \right) \left[
\delta \vec{g}\left( \vec{R}^{\left( 0\right) }\left( \vec{x},\vec{p},\tau
_{2}+\theta ,t_{0}\right) +\dfrac{\hbar \vec{k}}{2M_{a}}\left( T+\theta
\right) ,\tau _{2}+\theta \right) -\delta \vec{g}\left( \vec{R}^{\left(
0\right) }\left( \vec{x},\vec{p},\tau _{2}+\theta ,t_{0}\right) ,\tau
_{2}+\theta \right) \right] \right.  \notag \\
&&+\left. \theta \left[ \delta \vec{g}\left( \vec{R}^{\left( 0\right)
}\left( \vec{x},\vec{p},\tau _{1}+\theta ,t_{0}\right) +\dfrac{\hbar \vec{k}%
}{2M_{a}}\theta ,\tau _{1}+\theta \right) -\delta \vec{g}\left( \vec{R}%
^{\left( 0\right) }\left( \vec{x},\vec{p},\tau _{1}+\theta ,t_{0}\right)
,\tau _{1}+\theta \right) \right] \right\} .  \label{204}
\end{eqnarray}

\subsection{Part $\protect\phi _{Q}$}

Consider now $Q-$term (\ref{49b}). Since 
\begin{equation}
\left\{ 
\begin{array}{c}
\vec{R} \\ 
\vec{P}%
\end{array}%
\right\} \left( 
\begin{array}{c}
\vec{R}\left( \vec{R}\left( \vec{x},\vec{p},\tau _{1},t_{0}\right) ,\vec{P}%
\left( \vec{x},\vec{p},\tau _{1},t_{0}\right) +\hbar \vec{k}/2,\tau
_{3},\tau _{1}\right) , \\ 
\vec{P}\left( \vec{R}\left( \vec{x},\vec{p},\tau _{1},t_{0}\right) ,\vec{P}%
\left( \vec{x},\vec{p},\tau _{1},t_{0}\right) +\hbar \vec{k}/2,\tau
_{3},\tau _{1}\right) ,t,\tau _{3}%
\end{array}%
\right) =\left\{ 
\begin{array}{c}
\vec{R} \\ 
\vec{P}%
\end{array}%
\right\} \left( \vec{R}\left( \vec{x},\vec{p},\tau _{1},t_{0}\right) ,\vec{P}%
\left( \vec{x},\vec{p},\tau _{1},t_{0}\right) +\hbar \vec{k}/2,t,\tau
_{1}\right) ,  \label{49.1}
\end{equation}%
this term becomes%
\begin{eqnarray}
\phi _{Q}\left( \vec{x},\vec{p}\right) &=&-\dfrac{\hbar ^{2}}{24}\vec{k}_{u}%
\vec{k}_{v}\vec{k}_{w}\left\{ \int_{\tau _{1}}^{\tau _{2}}dt\chi
_{ikl}^{\prime }\left( \vec{\xi},t\right) \partial _{\vec{\pi}_{i}}\vec{R}%
_{u}\left( \vec{\xi},\vec{\pi},\tau _{1},t\right) \partial _{\vec{\pi}_{k}}%
\vec{R}_{v}\left( \vec{\xi},\vec{\pi},\tau _{1},t\right) \partial _{\vec{\pi}%
_{l}}\vec{R}_{w}\left( \vec{\xi},\vec{\pi},\tau _{1},t\right) \right.  \notag
\\
&&+\int_{\tau _{2}}^{\tau _{3}}dt\chi _{ikl}^{\prime }\left( \vec{\xi}%
,t\right) \left[ \partial _{\vec{\pi}_{i}}\left( \vec{R}_{u}\left( \vec{\xi},%
\vec{\pi},\tau _{1},t\right) -2\vec{R}_{u}\left( \vec{\xi},\vec{\pi},\tau
_{2},t\right) \right) \right] \left[ \partial _{\vec{\pi}_{k}}\left( \vec{R}%
_{v}\left( \vec{\xi},\vec{\pi},\tau _{1},t\right) -2\vec{R}_{v}\left( \vec{%
\xi},\vec{\pi},\tau _{2},t\right) \right) \right]  \notag \\
&&\times \left. \left[ \partial _{\vec{\pi}_{l}}\left( \vec{R}_{w}\left( 
\vec{\xi},\vec{\pi},\tau _{1},t\right) -2\vec{R}_{w}\left( \vec{\xi},\vec{\pi%
},\tau _{2},t\right) \right) \right] \right\} _{\left\{ 
\begin{array}{c}
_{\vec{\xi}} \\ 
_{\vec{\pi}}%
\end{array}%
\right\} =\left\{ 
\begin{array}{l}
_{\vec{R},} \\ 
_{\vec{P}}%
\end{array}%
\right\} \left( \vec{R}\left( \vec{x},\vec{p},\tau _{1},t_{0}\right) ,\vec{P}%
\left( \vec{x},\vec{p},\tau _{1},t_{0}\right) +\hbar \vec{k}/2,t,\tau
_{1}\right) }.  \label{49.2}
\end{eqnarray}%
When atoms move between pulses under the action of the homogeneous Earth
gravity field $\vec{g}$ and small but inhomogeneous perturbation $\delta 
\vec{g}(\vec{x},t)$ caused by the proof mass tensor (\ref{6b}) is caused
only by the proof mass, 
\begin{subequations}
\begin{eqnarray}
\chi _{ikl}^{\prime } &=&M_{a}\chi _{ikl},  \label{54a} \\
\chi _{ikl} &=&\partial _{x_{k}}\partial _{x_{l}}\delta g_{i}(\vec{x},t).
\label{54b}
\end{eqnarray}%
While we calculate the AI phase in the linear in proof mass gravity
approximation, it is sufficient to calculate the atom trajectory in Eq. (\ref%
{49.2}) in the absence of the proof mass, when 
\end{subequations}
\begin{equation}
\left\{ 
\begin{array}{c}
\vec{R} \\ 
\vec{P}%
\end{array}%
\right\} \left( \vec{x},\vec{p},t,t^{\prime }\right) =\left\{ 
\begin{array}{c}
\vec{R}^{\left( 0\right) } \\ 
\vec{P}^{\left( 0\right) }%
\end{array}%
\right\} \left( \vec{x},\vec{p},t,t^{\prime }\right)  \label{55}
\end{equation}%
and therefore 
\begin{equation}
\partial _{p_{i}}R_{j}\left( \vec{x},\vec{p},t,t^{\prime }\right) =\QDABOVE{%
1pt}{\delta _{ij}}{M_{a}}\left( t-t^{\prime }\right)  \label{56}
\end{equation}%
Using this result and Eqs. (\ref{16}, \ref{49.2})

\begin{eqnarray}
\phi _{Q}\left( \vec{x},\vec{p}\right)  &=&\dfrac{\hbar ^{2}}{24M_{a}^{2}}%
k_{i}k_{j}k_{l}\left[ \int_{\tau _{1}}^{\tau _{2}}dt\chi _{ijl}\left( \vec{%
\xi},t\right) \left( t-\tau _{1}\right) ^{3}\right.   \notag \\
&&\left. +\int_{\tau _{2}}^{\tau _{3}}dt\chi _{ijl}\left( \vec{\xi},t\right)
\left( \tau _{3}-t\right) ^{3}\right] _{\vec{\xi}=\vec{R}\left( \vec{R}%
\left( \vec{x},\vec{p},\tau _{1},t_{0}\right) ,\vec{P}\left( \vec{x},\vec{p}%
,\tau _{1},t_{0}\right) +\hbar \vec{k}/2,t,\tau _{1}\right) }  \label{57}
\end{eqnarray}%
Since we are interested in calculating of the $Q-$term to the 2nd order in
recoil momentum $\hbar \vec{k}$, we then can neglect recoil in the brackets
of Eq. (\ref{57}). We then apply the multiplication law (\ref{31}) and
obtain 
\begin{equation}
\phi _{Q}\left( \vec{x},\vec{p}\right) =\dfrac{\hbar ^{2}}{24M_{a}^{2}}%
k_{i}k_{j}k_{l}\left\{ \int_{\tau _{1}}^{\tau _{2}}dt\chi _{ijl}\left[ \vec{R%
}\left( \vec{x},\vec{p},t,t_{0}\right) ,t\right] \left( t-\tau _{1}\right)
^{3}+\int_{\tau _{2}}^{\tau _{3}}dt\chi _{ijl}\left[ \vec{R}\left( \vec{x},%
\vec{p},t,t_{0}\right) ,t\right] \left( \tau _{3}-t\right) ^{3}\right\} 
\label{58}
\end{equation}%
Substituting $t=\tau _{1}+\theta $ for the 1st term of Eq. (\ref{58}), and $%
t=\tau _{2}+\theta $ for the 2nd term one finds%
\begin{equation}
\phi _{Q}\left( \vec{x},\vec{p}\right) =\dfrac{\hbar ^{2}}{24M_{a}^{2}}%
k_{i}k_{j}k_{l}\int_{0}^{T}d\theta \left\{ \theta ^{3}\chi _{ijl}\left[ \vec{%
R}\left( \vec{x},\vec{p},\tau _{1}+\theta ,t_{0}\right) ,\tau _{1}+\theta %
\right] +\left( T-\theta \right) ^{3}\chi _{ijl}\left[ \vec{R}\left( \vec{x},%
\vec{p},\tau _{2}+\theta ,t_{0}\right) ,\tau _{2}+\theta \right] \right\} .
\label{59}
\end{equation}

\begin{acknowledgments}
\end{acknowledgments}

Author is appreciated to Drs. B. Young, S. Libby, M. Matthews, T. Loftus, M.
Shverdin, V. Sonnad and A. Zorn for fruitful discussion and collaboration.
Special gratefulness to Dr. A. Zorn for assistance in derivations of Eqs. (%
\ref{204}, \ref{59}).

This work was performed under the auspices of the Defense Threat Reduction
Agency by AOSense, Inc. under Contract HDTRA1-13-C-0047.


\begin{thebibliography}{9}
\bibitem{c1} B. Dubetsky and M. A. Kasevich, Phys. Rev. A \textbf{74},
023615 (2006).

\bibitem{c2} J. B. Fixler, G. T. Foster, J. M. McGuirk, M. A. Kasevich1,
Science \textbf{315}, 74 (2007).

\bibitem{c3} G. Rosi, F. Sorrentino, L. Cacciapuoti, M. Prevedelli \& G. M.
Tino, Nature \textbf{510}, 518 (2014).

\bibitem{c4} G. W. Biedermann, X. Wu, L. Deslauriers, S. Roy, C.
Mahadeswaraswamy, M. A. Kasevich, http://arxiv.org/abs/1412.3210.

\bibitem{c5} B. Dubetsky, Private communications, 2008

\bibitem{c6} B. Dubetsky, http://arxiv.org/abs/1407.7287, Eq. (3)
\end{thebibliography}
\end{document}